\documentclass[prd,twocolumn,aps,nofootinbib,nobibnotes,superscriptaddress,preprintnumbers]{revtex4-1}

\usepackage{appendix}
\usepackage{comment}
\usepackage{color}
\usepackage{amsmath}
\usepackage{amsfonts}
\usepackage{graphicx}
\usepackage[dvipsnames]{xcolor}
\definecolor{refs}{RGB}{245,156,74}
 \usepackage[colorlinks=true,hyperfootnotes=true,citecolor=cyan]{hyperref}
\usepackage{animate}

\newcommand{\rs}{r_{\rm s}}
\newcommand{\as}{a_{\rm s}}

\newcommand{\dd}{{\rm d}}

\newcommand{\Lag}{\mathcal{L}}

\newcommand{\mU}{\mathcal{U}}

\newcommand{\be}{\begin{equation}}
\newcommand{\ee}{\end{equation}}
\newcommand{\bea}{\begin{eqnarray}}
\newcommand{\eea}{\end{eqnarray}}
\renewcommand{\bf}[1]{{\textbf{#1}}}

\newcommand{\Fd}{\tilde{F}}
\newcommand{\mpl}{M_{\rm Pl}}
\newcommand{\mK}{{\mathcal{K}}}
\newcommand{\mM}{{\mathcal{M}}}
\newcommand{\tini}{t_\star}
\newcommand{\rini}{r_\star}
\newcommand{\Rdm}{R_{\rm DM}}
\newcommand{\Rb}{R_{\rm B}}

\newcommand{\rhode}{\rho_{\rm DE}}
\newcommand{\Lambdae}{\Lambda_{\rm e}}

\begin{document}

\title{Charged Dark Matter and the $H_0$ tension}

\author{Jose Beltr\'an Jim\'enez}
\email[]{jose.beltran@usal.es}
\affiliation{Departamento de F\'isica Fundamental and IUFFyM, Universidad de Salamanca, E-37008 Salamanca, Spain.}
\author{Dario Bettoni}
\email[]{bettoni@usal.es}
\affiliation{Departamento de F\'isica Fundamental and IUFFyM, Universidad de Salamanca, E-37008 Salamanca, Spain.}

\author{Philippe Brax}
\email[]{philippe.brax@ipht.fr}
\affiliation{Institut de Physique  Th\'eorique, Universit\'e Paris-Saclay,CEA, CNRS, F-91191 Gif-sur-Yvette Cedex, France.}

\begin{abstract}
We consider cosmological models where dark matter is universally charged under a dark Abelian gauge field. This new interaction is repulsive and competes with gravity on large scales and in the dynamics of galaxies and clusters. We focus on non-linear models of dark electrodynamics where the effects of the new force are screened within a K-mouflage radius that helps avoiding traditional constraints on charged dark matter models. We discuss the background cosmology of these models in a Newtonian approach and show the equivalence with relativistic Lema\^itre models where an inhomogeneous pressure due to the electrostatic interaction is present. In particular, after foliating the Universe using  spherical shells, we find that dark matter shells with initially different radii do not evolve similarly as they exit their K-mouflage radii at different times, resulting in a breaking of the initial comoving evolution. In the large time regime, the background cosmology is described by a comoving but inhomogeneous model with  a reduced gravitational Newton constant and a negative curvature originating from the electrostatic pressure. In this model, baryons do not directly feel the electrostatic interaction, but are influenced by the inhomogeneous matter distribution induced by the electric force. We find that shells of smaller radii evolve faster than the outer shells which feel the repulsive interaction earlier. This mimics the discrepancy between the large scale Hubble rate and the local one. Similarly, as  galaxies and clusters are not screened by the new interaction, large scale global flows would result from the existence of the new dark electromagnetic interaction.  
\end{abstract}
\maketitle
\tableofcontents

\section{Introduction}

The uniqueness of General Relativity as the low energy effective field theory for a massless spin-2 particle respecting local Lorentz invariance and locality is at the heart of the ubiquitous presence of additional degrees of freedom in infrared modifications of gravity. In relation with the problem of dark energy, theories featuring new scalar fields are especially appealing and, consequently, their phenomenological consequences have been extensively explored. One of the most interesting features of these scalar fields, that typically mediate fifth forces, is the presence of screening mechanisms that allow to evade local gravity tests. It is also the reason why these fields exhibit an elusive character. There is a variety of screening mechanisms \cite{Khoury:2010xi} that can be classified according to the type of operators that drive them, namely: Chameleon \cite{Khoury:2003aq,Khoury:2003rn}, symmetron \cite{Pietroni:2005pv,Olive:2007aj,Hinterbichler:2010es}, dilaton \cite{Brax:2010gi} if the screening relies on non-derivative operators, K-mouflage/Kinetic screening \cite{Babichev:2009ee,Brax:2012jr,Brax:2014wla,Brax:2014yla} if the mechanisms originate from operators with first derivatives and Vainshtein \cite{Vainshtein:1972sx,Babichev:2013usa} if the relevant operators for the screening contain second derivatives. 

In view of the rich phenomenology provided by scalar field theories featuring screening mechanisms, considering that the three fundamental interactions present in the standard model (other than gravity) are mediated by gauge bosons and that the Universe contains a dark sector where similar types of gauge interactions to those of the visible sector could be expected, it is certainly  alluring to envisage the existence and phenomenology of screening mechanisms for spin-1 fields. In particular, we will be interested in massless spin-1 fields (for the massive case see e.g. \cite{BeltranJimenez:2013fca}). In this scenario, the absence of Galileon-type  interactions \cite{Deffayet:2013tca} makes it natural to consider a screening \`a la $K$-mouflage. As a matter of fact, this idea was already realised in the Born-Infeld electromagnetism that can arguably be considered as the first screening mechanism of this type, although with a different aim \cite{Born:1933pep,Born:1934gh}. In this work, we will explore some consequences of having a dark U(1) gauge boson that mediates an extra force following the approach of \cite{BeltranJimenez:2020csl}, sharing similarities with  earlier works \cite{Kaloper:2009nc}. This scenario can have an important impact on the cosmological evolution of the Universe. Let us assume that the gauge boson mediates a dark interaction for the dark matter particles and that the early Universe underwent a phase of {\it dark matter genesis} where only one type of charged DM particles survived. In the following, we will be agnostic about the mechanism which could realise this separation between the particles of dark matter and their antiparticles but mechanisms that realise such behaviour have been put forward \cite{Goolsby-Cole:2015chd}.
 We will focus on the consequences of the presence of a charged dark matter species leaving the detailed study of possible mechanisms for its production to further work. 
Let us notice, however, that mechanisms that generate asymmetry in the dark matter sector have been proposed  \cite{Petraki:2013wwa} and that cosmological models in which a net electric charge is present have been considered, see for example \cite{Soriano:2019cmu} and references therein. As we will see, in the early Universe, the screening radius is larger than the Hubble horizon and, therefore, the DM component is impervious to the dark boson interaction. As the Universe expands, the horizon grows until, at some point, it becomes larger than the largest screening radius hence allowing astrophysical objects to feel the repulsive force, if certain conditions are met. In particular, during the early stage of horizon screening, the DM component is subject to gravitational collapse and it forms halos as in the standard model. However, in our scenario, each halo will have its own screening radius so that DM halos separated by distances larger than twice the screening scale will feel an additional repulsive force mediated by the gauge boson as soon as the horizon crosses the screening scale. The repulsive nature of the force, as opposed to the usual attractive force produced by scalar fields, is of course due to the spin-1 nature of the gauge boson. This can have two effects. The first one is at the cosmological background level. We will show that below a certain redshift the extra electromagnetic interaction acts to lower Newton's constant and adds a contribution to the spatial curvature. As we will argue, the appearance of the low redshift dark electric repulsion can shed some light on the pressing tension for the value of $H_0$ as measured locally \cite{Riess:2019cxk,Wong:2019kwg,Birrer:2018vtm} and inferred from Cosmic Microwaves Background (CMB) anisotropies observations \cite{Aghanim:2018eyx} (see \cite{Verde:2019ivm} for a recent state of the art summary). Typically we expect that for couplings of dark matter to the dark electromagnetic interaction of order one as compared to gravity, i.e. $\beta=\mathcal O(1)$, and for a suppression scale $\Lambdae$ of the order of the dark energy scale, the large scale value of the Hubble rate can be valid down to redshifts of order $0.5$ allowing BAO measurements (Baryon Acoustic Oscillations) to corroborate the CMB results whilst the local value of the Hubble rate can be different by a factor of the order of ten percent. In a similar vein, once galaxy clusters have formed, their dark charges can imply that the peculiar velocities of both the clusters themselves and the galaxies which are embedded within them are modified by the new repulsive electromagnetic interaction. This should leave an imprint on the late time distribution of structures on large scales. The existence of a long-range interaction for the dark matter sector due to a dark U(1) charge has been extensively considered in the literature and their observational signatures analysed, especially as an attempt to alleviate the structure formation problems of the standard collisionless dark matter paradigm \cite{Gradwohl:1992ue,Ackerman:mha,ArkaniHamed:2008qn,Feng:2009mn,Feng:2009hw,Aarssen:2012fx,Garny:2018grs}. While such models can provide promising mechanisms to explain the small scale anomalies of the standard model, the additional long-range interaction is also tightly constrained (see e.g. \cite{Agrawal:2016quu} and references therein). Our scenario however crucially differs from these models precisely in the existence of non-linearities in the dark gauge sector that suppress the coupling constant both at high redshift (thus avoiding effects on the freeze-out time or the core of the structure formation period) and on small scales at present times. As discussed above and in more detail below, observational signatures are however expected in an interesting range of scales.

In this article we will study the cosmological evolution of the Universe in the presence of the additional dark electromagnetic force. As a first step, we will focus on the background cosmology of the Universe when this new dark force is present. In the absence of the dark interaction, the background evolution of the Universe can be appropriately described using the {\it cosmological principle} which posits that the Universe is homogeneous and isotropic at sufficiently large scales. As well known, this results in the usual relativistic description of the dynamics of the Universe using the Friedmann-Lema\^itre-Robertson-Walker (FLRW) metric. In this setting, the Universe can be foliated by spherical sections thanks to isotropy. Each of these sections follow co-moving trajectories described by a common scale factor. 
Notice that this background dynamics can be fully described in terms of a Newtonian approach, see for example \cite{Gibbons:2013msa}. When the dark force is introduced, {\it isotropy} of the background cosmology is maintained whilst {\it homogeneity} is lost. This follows from the scale dependence of the additional interaction which is non-linear and breaks scale invariance. As a result, different spherical sections of the background cosmological description of the Universe on large scales will not evolve in a co-moving fashion. A fully relativistic treatment of this system can be given and necessitates the use of Lema\^itre metrics \cite{Lemaitre:1933gd} generalising the more common Lema\^itre-Tolman-Bondi approach \cite{Lemaitre:1933gd,Tolman:1939jz,Bondi:1947fta} to inhomogeneous space-times. The difference between the two metrics springing from the gradient of pressure coming from the presence of a radial electric field.\footnote{In the isotropic situations relevant for us the magnetic field plays no role since an expanding charged sphere does not generate any magnetic field. At the perturbation level, when going beyond the background cosmology as described here, magnetic fields will be generated by the peculiar velocities of cosmological structures. The resulting magnetic field depends linearly on the velocity field $v$ and the resulting Lorentz force is $\mathcal{O}(v^2)$. As dark matter velocities are always small of order $10^{-3}$, the magnetic interaction induced by the relative motion of charges is suppressed compared to the electric interaction. As a result, only taking into account the electric interaction, as in the isotropic description of the background cosmology we give here, is not thwarted by the magnetic effects due to local inhomogeneities.} In section \ref{Sec:Lemaitre} we sketch such a relativistic treatment by introducing a dark fluid mimicking the effects of the electric field. The full relativistic case where such a dark fluid follows from non-linear electrodynamics has been derived in \cite{BeltranJimenez:2021imo}. As in the FRLW case, we will tackle the background cosmological evolution by resorting to a Newtonian cosmology approach where the Universe is dust dominated and the particles are subject to both gravity and the new electric force. We will explicitly see how the evolution of the spherical sections of the Universe in the presence of the dark electric force, starting from an initial state where the Universe evolves according to the standard Hubble flow at high redshift, enters a phase where the comoving motion ceases and different scales expand at different rates as they exit their corresponding screening radii. Asymptotically, however, the Universe recovers a comoving evolution where all the scales expand at the same rate again, but in the process an inhomogeneous density profile is generated. We will see how these results can be interpreted in terms of inhomogeneous models with spherical symmetry such as Lema\^itre models. We will confirm our results by numerically solving the Newtonian evolution for a discrete set of shells that will allow us to address the crucial issue of a regular stream flow without shell crossing and also analysing the effects of adding an uncharged baryonic component and a cosmological constant.

The background cosmological description of the Universe in terms of an  inhomogeneous and isotropic Universe allows us to suggest that the present $H_0$ tension may eventually be resolved by taking into account such dynamical inhomogeneities in the dark matter distribution induced by the dark force. Notice that these dynamical inhomogeneities differ greatly from the known inhomogeneities in the baryonic distribution, obtained because we are located in an underdense region, which can be taken into account using a Lema\^itre-Tolman-Bondi description of the Universe. The effect of these local inhomogeneities is known to be twenty times too small to close the $H_0$ tension \cite{DiValentino:2020zio,Wu:2017fpr,Kenworthy:2019qwq}. The inhomogeneities that we obtain here would affect the local dynamics of dark matter as the dark force would not be screened compared to the screening of its effects early in the Universe. A precise comparison of this proposal to observations is left for future work. 

The article is organised as follows: In Section \ref{Sec:nonlinearEM} we will briefly describe some general properties of non-linear electromagnetism, emphasising the physical origin of the screening mechanism. In Section \ref{Sec:Newtonian} we study the Newtonian cosmology, with a discussion on its validity, how an effective Friedmann equation is obtained and its relation to inhomogeneous Lema\^itre models. We confirm our analytical results by numerically solving the system in Section \ref{Sec:numerics} for DM only and including baryons which are taken to be decoupled from dark electromagnetism. We also discuss the shell-crossing condition. In this section we show how the additional electric repulsion can help to alleviate the $H_0$ tension. Finally, we discuss some observational consequences in Section \ref{Sec:Discussion}.

\bf{Conventions:} The field strength of the gauge field is $F_{\mu\nu}=\partial_\mu A_\nu-\partial_\nu A_\mu$. Its dual is defined as $\Fd^{\mu\nu}=\frac12 \epsilon^{\mu\nu\alpha\beta} F_{\mu\nu}$. The electric and magnetic components are $E_i=F_{0i}$ and $B_i=\Fd_{0i}$. We will work with mostly plus signature for the metric.

\section{Non-linear electrodynamics}\label{Sec:nonlinearEM}

\subsection{The non-linear U(1) model}
The properties of non-linear electrodynamics have been extensively considered in the literature \cite{Plebanski:1970zz,Boillat:1970gw} so here we will only give the most relevant aspects for our purposes. Let us consider a theory for an Abelian gauge spin-1 field $A_\mu$ described by the Lagrangian
\be
\Lag=\mK(Y,Z)+ J^\mu A_\mu
\ee
with $\mK$ and arbitrary function of $Y=-\frac14 F_{\mu\nu} F^{\mu\nu}$ and $Z=-\frac14 F_{\mu\nu} \Fd^{\mu\nu}$ and $J^\mu$ the conserved current that describes the charged sector. Notice that the coupling to charged matter is not modified and corresponds to the one of linear electrodynamics. The specific nature of the DM charged sector is not crucial for our purposes here and it will suffice to assume that they behave like charged point-like particles\footnote{The precise scenario for the DM charged sector could be relevant for studies of direct or indirect detection where the precise couplings to the Standard Model particles are crucial.}. In terms of the electric and magnetic components we have $Y=\frac12(\vec{E}^2-\vec{B}^2)$ and $Z=\vec{E}\cdot \vec{B}$. The non-linear dependence on $Y$ and $Z$ will become relevant at  some scale $\Lambdae$ that will control the classical non-linearities. Since quantum corrections are expected to enter with derivatives of the field strength\footnote{We refer here to quantum corrections due to self-interactions of the vector field. The coupling to matter fields will generate quantum corrections to $\mK$ that are suppressed by the mass of the particle running in the internal loop. A paradigmatic example is the Euler-Heisenberg Lagrangian. Corrections due to self-interactions arise in a non-trivial background as one expands the quantum fluctuations around this field configuration. The quantum corrections depend on the background. In the case of scalar field theory like K-mouflage with a shift symmetry, it has been shown that these quantum corrections appear only as higher order derivative corrections and therefore leave the original Lagrangian non-renormalized \cite{Brax:2016jjt}.} $\partial^\ell F^n$, there should be a regime where classical non-linearities are relevant and within the regime of validity of the effective field theory (EFT). In this regime we can have $F_{\mu\nu}\sim\Lambdae^2$ as long as $\partial\ll\Lambdae$. 

The field equations for the gauge field are
\be
\nabla_\nu\Big(\mK_Y F^{\mu\nu}+\mK_Z \Fd^{\mu\nu}\Big)=J^\mu
\ee
where the current $J^\mu$ acts as its source. As usual, these equations can be complemented by the Bianchi identities $\nabla_\mu\Fd^{\mu\nu}\equiv 0$. We will consider now a static source with $J^\nu=(\rho_q,\vec{0})$. The field equations in this situation reduce to
\be
\nabla\cdot\Big(\mK_Y\vec{E}+\mK_Z\vec{B}\Big)=\rho_q,\quad\quad
\nabla\times\Big(\mK_Y\vec{B}+\mK_Z\vec{E}\Big)=0\,.
\ee
Assuming parity invariance, it is non-contradictory to consider a vanishing magnetic field. Indeed, parity invariance imposes a $\mathbb{Z}_2$ symmetry with respect to $Z$, i.e., the Lagrangian can only depend on $Z^2$. In that case $\mK_Z=2Z\partial \mK/\partial Z^2$ that vanishes identically for $\vec{B}=0$ provided $\mK$ is an analytical function. The equations then read
\be
\nabla\cdot\Big(\mK_Y\vec{E}\Big)=\rho_q\,.
\ee
If the source is spherically symmetric, we can integrate over a sphere enclosing the source so that Gauss' theorem gives
\be
\mK_Y\vec{E}=\frac{Q}{4\pi r^3}\vec{r}
\label{eq:solE}
\ee
with $Q=\int\rho_q\dd^3x$ the total charge inside the sphere.
This is the usual result obtained in classical electrodynamics dressed by the extra $\mK_Y$ factor. As for scalar K-mouflage models, the non-linearities and the origin of screening lie in the fact that $\mK_Y\ne 1$ and can become very large.

 We will embed this model of nonlinear electrodynamics in the cosmological  description of the Universe. Typically, dark matter will be charged under the new $U(1)$ interaction and at the background cosmological level that we consider here, we will resort to an isotropic  and inhomogeneous space-time as the presence of the non-linear electromagnetic interaction breaks scale- invariance by introducing an intrinsic scale $\Lambdae$ in the model. On the other hand, it is exactly this non-linear character of the interaction that allows for the presence of a screening phenomenon that we describe next.

\subsection{Screening} 

The screening mechanism is now trivial to understand. If $\mK$ is an analytic function of $Y$ such that $\mK(Y)\sim Y$ for $Y\rightarrow 0$ we have that
\be
E\simeq\frac{Q}{4\pi r^2},\quad\quad r\rightarrow \infty\,.
\ee
As we approach the object, the electric field grows and the non-linearities become more relevant. Since the non-linear terms are controlled by the scale $\Lambdae$, we can expect them to become relevant when $E\sim \Lambdae^2$ that occurs at a scale parameterically given by
\be
\frac{Q}{4\pi r^2\Lambdae^2}=1\Rightarrow\rs=\kappa\sqrt{\frac{Q}{4\pi}}\Lambdae^{-1}
\ee
where $\kappa$ is a number (typically of order unity) that depends on the specific theory under consideration. Below this scale, the electric field is given by
\be
E\simeq\frac{Q}{4\pi\mK_Y r^2},\quad\quad r\ll \rs
\ee
where we clearly see the screening at work, i.e. it is  induced by having a large $\mK_Y$ that suppresses the electric field. This can be interpreted as a screening of the effective charge $Q$ that is classically re-dressed by the electric field.

If we have a given distribution of particles of mass $m$ and charge $q$, then the dominant monopole electric field (which is the only contribution for a spherical distribution) can be expressed in terms of the mass of the distribution. If we denote by \footnote{The normalisation is introduced for convenience so that $\beta$ directly measures the relative strength of gravity and the electric force outside the screening radius. Notice the factor of 2. }
\begin{equation}
\beta\equiv \frac{\sqrt{2}Q\mpl}{M}
\end{equation}
the charge-to-mass ratio of the object, then  we can express the electric field as
\be
\vec{E}=\beta\frac{M}{4\sqrt{2}\pi \mpl \mK_Y r^3}\vec{r}\,.
\label{eq:solEM}
\ee
On the other hand, the screening radius can also be expressed in terms of the mass as
\be\label{eq:r_s}
\rs=\lambda\sqrt{\frac{M}{\mpl}}\Lambdae^{-1},\quad\text{ with}\quad\lambda\equiv\kappa\sqrt{\frac{\beta}{2\sqrt{2}\pi}}\,.
\ee
Before proceeding further with the analysis, let us anticipate some order of magnitude estimates for the model's parameters. In the following we will be interested in the Hubble tension and the role that the new dark electromagnetic interaction could be in alleviating this discrepancy. In particular, we will see that taking $\beta={\mathcal O}(1)$ and for a scale $\Lambdae$ close to the dark energy scale the local Hubble rate can be made compatible with a ten percent difference with the Hubble rate on large scales. These values of $\beta$ and $\Lambdae$ will be the typical templates for our model. 
If we take $\lambda=\mathcal{O}(1)$ in equation \eqref{eq:r_s} , which is equivalent to having $\beta\sim \mathcal O (1)$, the screening radius of a particle of mass $m=\mathcal{O}$(GeV) is $\rs\simeq 10^{-9}\Lambdae^{-1}$. If we now use the dark energy scale for $\Lambdae$, i.e., $\Lambdae=\sqrt{\mpl H_0}\simeq 10^{-3}$eV, we obtain $\rs\sim 10^{-6}$eV$^{-1}\sim 10^{-13}$m. This is a very small screening radius implying that in practice particles act as unscreened point-particle objects. However, due to the non-linearities of the theory, large accumulations of mass such as galaxies or cluster do not act as the superimposition of individual particles, i.e. collective effects take place, and they will be screened. 

In the presence of a massive object of mass $M$, a test particle of mass $m$ and charge $q$ will  experience a force due to gravity and the electric field that can be written as
\be
\vec{F}=-G\frac{mM}{r^3}\vec{r}+\frac{qQ}{4\pi\mK_Y r^3}\vec{r}=-G\frac{mM}{r^3}\left(1-\frac{\beta^2}{\mK_Y}\right)\vec{r}\,.
\ee
This expression clearly shows how at large distances where $\mK_Y\simeq 1$, the electric force contributes with a strength $\beta^2$ relative to gravity while inside the screening radius where $\mK_Y\gg1$ the electric force is strongly suppressed. This screening phenomenon shows similarities with other screenings appearing for scalar interactions such as the Vainshtein and K-mouflage phenomena. Inside the screening radius, the interaction with a point particle is shielded. Outside the screening radius, the effacement theorem, see appendix \ref{app:A}, applies and a macroscopic object acts under the dark electromagnetic interaction as a point particle. This implies that two macroscopic objects separated by a distance larger than the sum of their screening radii interact with the unscreened dark interaction of strength $\beta^2$. We will see in section \ref{Sec:Discussion} that this could have interesting phenomenological consequences for the dynamics of galaxy clusters.  
Let us also notice  that the additional interaction effectively weakens gravity due to the electric repulsion outside the screening radius, in high contrast to the scalar field theories. In fact, we can encapsulate the effect of the electric force into an effective scale-dependent Newton's constant
\be
\frac{G_{\rm eff}}{G}=1-\frac{\beta^2}{\mK_Y}\,.
\label{supp}
\ee
This is one of the main effects that we will exploit in this work. Since the absence of ghost-like perturbations requires $\mK_Y>0$, the effective Newton's constant is always reduced for stable theories. Notice that taking $\beta={\cal O}(1)$ could result in the dark interaction overcoming the gravitational force. This would lead to an "anti"-gravity effect between particles subject to this new interaction. Of course, the well-known constraints on gravitational interactions between baryons imply that this behaviour is excluded. On the other hand, a large and repulsive interaction between unscreened objects made out of dark matter  can be made plausible as we explain in section \ref{Sec:Discussion} where phenomenological constraints are applied.

Before proceeding to the main core and for illustrative purposes, let us briefly give some details for two paradigmatic non-linear electrodynamics.

\subsection{Two examples}

\bf{Born-Infeld electromagnetism.} 
As commented in the introduction, this is allegedly the first electromagnetic theory ever exploiting  screening properties. The Lagrangian can be written in the following two alternative forms:
\begin{widetext}
\be
\Lag_{\rm BI}=\Lambdae^4\left[1-\sqrt{-\det\left(\eta_{\mu\nu}+\frac{1}{\Lambdae^2}F_{\mu\nu}\right)}\right]=\Lambdae^4\left[1-\sqrt{1-\frac{2Y}{\Lambdae^4}-\left(\frac{Z}{\Lambdae^4}\right)^2\;}\right]\,.
\ee
\end{widetext}
The electric field for this Lagrangian can be solved analytically and is given by

\be
E=\frac{1}{\sqrt{1+\left(\frac{\rs}{r}\right)^4}}\frac{Q}{4\pi r^2},\quad\text{with}\quad\rs=\sqrt{\frac{Q}{4\pi}}\Lambdae^{-1}\,.
\ee
Notice that this is the unique solution and that several branches do not appear contrary to what happens in the power law example provided in the following section.
Clearly, for $r\gg\rs$, the electric field approaches the Maxwellian solution, while at short distances $r\ll\rs$, the electric field is 
\be
E\simeq \left(\frac{r}{\rs}\right)^2\frac{Q}{4\pi r^2}=\Lambdae^2
\ee
so its value saturates to an upper constant bound, as the theory was designed for. In particular, the electric field does not diverge at the origin and this regularises the classical self-energy of point-like particles.
A summary of the properties of the Born-Infeld case can be found in table \ref{Table:examples}. 

\bf{Power law correction.} Another useful example of non-linear electrodynamics is adding a power law correction to the Maxwellian Lagrangian
\be
\Lag_{n}=Y+\Lambdae^4\left(\frac{Y}{\Lambdae^4}\right)^n
\ee
with $n$ a dimensionless parameter that must be $n>1$ in order to recover Maxwell electromagnetism at large distances (small electromagnetic fields). We could add contributions depending on $Z^2$ as well, but since these trivialise for static purely electric configurations as the ones we consider here, they are not relevant. They should be relevant however for the behaviour of the perturbations. The equation for the electric field can be written as
\be
\left[1+n\left(\frac{E^2}{2\Lambdae^4}\right)^{n-1}\right]E=\frac{Q}{4\pi r^2}\,.
\ee
This equation exemplifies the expected feature that non-linear electromagnetism exhibit several branches, one of which is continuously connected to the Maxwell solution at infinity. This is the one we will be interested in. The case of Born-Infeld is also special due to the absence of multi-branching.  The screening radius can be computed as
\be
n\left(\frac{E^2}{2\Lambdae^4}\right)^{n-1}=1\quad\Rightarrow\quad\rs=\left[2n^{1/(1-n)}\right]^{-1/4}\sqrt{\frac{Q}{4\pi}}\Lambdae^{-1}.
\ee
again in accordance with the general expression \eqref{eq:r_s} so it is parameterically determined by the non-linear scale $\Lambdae$. In the present case of a polynomial equation of degree $2n-1$ the solution can be found by solving the associated algebraic equation.  Since a general  algebraic  solution  is more involved to obtain, and generically does not exist for $n\ge 3$, we will limit ourselves to computing the behaviour of the electric field below the screening scale
\be
E\simeq \kappa_n \left(\frac{r}{\rs}\right)^{\frac{2}{1-2n}} \Lambdae^2
\ee
with $\kappa_n=\sqrt{2n^{1/(1-n)}}$. This expression clearly shows that the electric field is suppressed with respect to its Maxwellian counterpart provided $n>1$. Notice however that the electric field is  divergent at the origin. A summary of the properties of the quadratic ($n=2$) case can be found in table \ref{Table:examples}.

\begin{widetext}
\begin{center}
\begin{table}[!htbp]
\begin{tabular}{|c|c|c|c|}
 Theory & Lagrangian & Function & Screening scale \\ \hline\hline
   & & &\\
 Born-Infeld &$\frac{\mathcal{L}_{\rm BI}}{\Lambdae^4}=1-\sqrt{-\det\big(\eta_{\mu\nu}+\frac{1}{\Lambdae^2}F_{\mu\nu}\big)}$  & $F(x)=\frac{1}{\sqrt{1+x^{-4}}}$&$\rs=\frac{1}{\Lambdae}\sqrt{\frac{Q}{4\pi}}$ \\&&&\\ 
 &&&\\
 Quadratic ($n=2$) & $\mathcal{L}_2 =-\frac{1}{4}F_{\mu\nu} F^{\mu\nu}+\left(\frac{F_{\mu\nu}F^{\mu\nu}}{4\Lambdae^4}\right)^2$ & $\left[1+\frac{F(x)}{x^4}\right]F(x)=1$   &$\rs=\frac{1}{\Lambdae}\sqrt{\frac{Q}{4\pi}}$ \\&&&\\\hline
\end{tabular}
\vspace{0.1cm}
 \caption{We give two specific examples of non-linear electromagnetism featuring a screening mechanism. Accidentally, the screening scale coincides for these two theories, but the parametric scaling with $Q$ and $\Lambdae$ is universal. We also have introduced, for future convenience, the function $F(x)$ that incorporates the non-linear properties of each specific model.}\label{Table:examples}
\end{table}
\end{center}
\end{widetext}

\section{Newtonian Cosmology}\label{Sec:Newtonian}
\subsection{Formalism}
We are interested in studying a Universe with a dark electromagnetic interaction described by the non-linear theories as introduced above and  featuring a screening mechanism. We will commence our analysis by studying the Newtonian cosmology within this scenario. A nice discussion of Newtonian cosmology can be found in \cite{HARRISON1965437}. Here we will content ourselves with highlighting the most relevant points for our purposes. For the sake of simplicity, we will assume a Universe filled with an ensemble of non-relativistic massive particles that interact through the dark electromagnetic force in addition to the usual gravitational attraction. Furthermore, we will assume that all the particles have the same mass and charge. This could be the case if we only consider the dynamics of dark matter (DM) particles and we assume them to be conformed by one single species that has a certain dark charge. We will extend our analysis by including an uncharged baryon component later.

The idea is then to consider an infinite distribution of particles of density $\rho$ which we will assume to be initially homogeneous (for a detailed account on discrete cosmology see \cite{Gibbons:2013msa}). A little digression on the problem of  infinite distributions seems in order. If the particles interact via a long-range force, dynamics are not well-posed due to the divergence of the force. The Newtonian potential is precisely on the verge of being convergent. This follows from the fact  that, assuming a constant density profile, the force at a point $\vec{x}$ is formally given by 
\be
\vec{F}\propto \int\dd^3x'\frac{\vec{x}-\vec{x}'}{\vert\vec{x}-\vec{x}'\vert^3}\,.
\ee
The integrand for large distances goes as a constant and, therefore, the force is linearly divergent so we can assign to it any value  by simply performing an appropriate arrangement of the integration volume. For instance, if the integration is performed symmetrically with respect to $\vec{x}$ we will obtain $\vec{F}=0$. More technically, while the integral can be made convergent, it is not conditionally convergent and we need some physical guidance to give it some physical sense. The way to treat this problem is by suitably defining the infinite problem. For that, we can  take a sphere of a given radius $R$ and only at the end do we take the limit $R\rightarrow\infty$. Certainly, the ill-defined final limit introduces a dependence on the initial geometry we start with. Since we want to achieve a spherically symmetric solution, a sphere is the appropriate initial geometry. Another approach that would suffice for our purposes here would be to consider actually an isotropic distribution of  matter inside a sphere of a radius $R$ much larger than the scales we are interested in, so the distribution would appear homogeneous and isotropic for our relevant observers. This would avoid taking the ill-defined limit, but it could introduce boundary effects. However, these will be negligible provided we work well inside the distribution.

Having clarified our approach to the Newtonian cosmology, we can proceed with our analysis. Let us consider the spherically symmetric shell distributions discussed above and describe their evolution with the time-dependent radial coordinate $R(t)$. We will assume that at some initial time $\tini$ the shells have initial positions $\rini=R(\tini)$ and velocities $v_\star=\dot{R}(\tini)>0$ as it corresponds to an initially expanding regime. The initial density profile will also be assumed to be isotropic so $\rho_\star=\rho(r_\star)$. Since the initial distribution of velocities is assumed to be radial, the initial spherical symmetry will be maintained throughout the evolution. Now we can follow such an evolution by using either Eulerian coordinates $(t,R(t))$ or Lagrangian coordinates $(t,\rini)$, i.e., we can either follow the evolution of each shell or follow the evolution of the density field. In terms of Eulerian coordinates, the dynamics is governed by the equations
\be
\ddot{R}(t,\rini)=-\frac{GM(R)}{R^2}\Big[1-\beta^2 F(R/\rs)\Big]\,.
\label{eq:dyn}
\ee
with 
\be
M(R)=4\pi\int_{y<R}\rho(y) y^2\dd y
\ee
the mass enclosed by a sphere of radius $R$. For the sake of generality, we have introduced the function $F(x)$ as a phenomenological parameterisation of the screening so that $F(x\gg1)\simeq 1$ and $F(x\ll1)\ll1$. The specific shape of this interpolating function depends on the precise theory and is given by $\mK_Y^{-1}$ expressed as a function of the radius. However, for our general arguments in the following, the detailed form of $F(x)$ is not needed. Having moving charged shells, one may object that magnetic forces should also be included. However, given the preserved spherical symmetry and that expanding charged spheres do not generate magnetic fields, the Lorentz force is in this case strictly zero. Furthermore, let us mention that magnetic forces that could be generated by peculiar motions due to deviations from spherical symmetry would be a next-to-leading order effect and can be safely ignored against the electric forces just as much as we neglect gravito-magnetic forces in the gravitational sector. 

For our purposes and to have a more direct connection to FLRW (Friedmann-Lema\^itre-Robertson-Walker), Lema\^itre  and LTB (Lema\^itre-Tolman-Bondi) models that we will present  in Section \ref{Sec:Lemaitre}, it is more convenient to use Lagrangian coordinates so that each shell is described by its radius at a given time, say $\rini=R(\tini)$. Using these coordinates amounts to foliating the spatial sections with the initial position of the shells. In that case, we can introduce the local scale factor defined by $a(t,\rini)\equiv R/\rini$ and rewrite the equations as
\be
\ddot{a}(t,\rini)=-\frac{G\mu(t,\rini)}{a^2(t,\rini)}\Big[1-\beta^2 F\Big(\small{\frac{a}{\as}}\Big)\Big]
\label{Eq:appgeneral}
\ee
with $\mu\equiv M\rini^{-3}$. Since the screening radius is $\rs=\lambda\Lambdae^{-1}\sqrt{M/\mpl}$, we have that $\as=\lambda\Lambdae^{-1}\sqrt{\mu\rini/\mpl}$. Now, $a$ must be considered as a function of time and the Lagrangian coordinate $\rini$. To simplify the notation, we will  denote  by $r\equiv \rini$ so $a=a(t,r)$ and $r$ is the (time-independent) radial coordinate. The mass density parameter $\mu$ can be written in terms of the scale factor as
\be
\mu(t,r)=4\pi \int_{\tilde{a}<a} \tilde{a}^2\dd\tilde{a} \rho(t,\tilde{a} r)\,.
\ee
If the evolution of the system is such that the different shells do not cross, this integral does not depend on time, so we can compute it by evaluating at $t=\tini$
\be
\mu(t,r)=\mu(\tini,r)=4\pi \int_{\tilde{a}<1} \tilde{a}^2\dd\tilde{a} \rho(\tini,\tilde{a}r)\,.
\ee
where we have used that $a_\star\equiv a(\tini,r)=1$. This simply reflects the fact that if there is no shell-crossing, the mass within a given shell is conserved and, consequently, it is determined by the initial configuration. Furthermore, if the initial density profile is uniform $\rho(\tini,\tilde{a}r)=\rho_\star$, we have
\be
\mu_\star=\frac{4\pi \rho_\star}{3}
\ee
so it is just a constant. In that case, the evolution equation becomes
\be
\ddot{a}=-\frac{4\pi G\rho_\star}{3a^2}\Big[1-\beta^2 F(a/\as)\Big]\,.
\label{eq:app}
\ee
with the screening scale factor given by
\be
\as(r)=\frac{\lambda}{\Lambdae}\sqrt{\frac{4\pi \rho_\star}{3\mpl} r}\,. 
\ee
If the non-linearities in the electric force sector were absent (i.e., $F(x)=1$ at all scales), the obtained evolution equation would not depend on $r$, indicating that all the shells evolve in the same way. The usual comoving motion of the particles would still be valid, though with a corrected Newton's constant accounting for the extra electric repulsion. In our case, however, we clearly see how the non-linear term breaks the self-similar evolution via the dependence on $r$ hidden in the screening scale. This means that we will have a comoving motion of the shells until they exit their respective screening radii. Since, in the absence of shell crossing, the screening radii scale as $\as\propto\sqrt{r}$, the more internal shells exit the screened regime earlier than the external shells\footnote{The presence of shell crossing complicates things in a substantial manner. For instance the screening scale could also depend on time. We will comment on shell crossing effects in Sections \ref{Sec:shellcrossing} and \ref{Sec:baryons}.}. This can potentially induce shell crossing since once a shell exits its screening scale it tends to expand faster due to the extra electric repulsion. For the moment, we will assume that no shell-crossing occurs in order to simplify the analysis, but we will come back to this point in Section \ref{Sec:shellcrossing}. Let us emphasise that the breaking of self-similarity in the evolution does not directly imply the appearance of self-crossing as we will corroborate with our numerical analysis in Section \ref{Sec:numerics}.

\subsection{Effective Friedmann equation}
From the evolution equations for the shells \eqref{eq:app}, we can obtain in a simple manner the effective Friedmann equation derived from Newtonian cosmology. For that,  we notice that it is possible to find a first integral of \eqref{eq:app} by means of the corresponding energy function, which is given by
\be
E(r)=\frac12\dot{a}^2-\frac{4\pi G\rho_\star}{3 a}+\mU(a)
\label{Eq:Energy1}
\ee
with
\be
\mU(a)\equiv\mU_{\star}-\frac{4\pi G\rho_\star}{3}\beta^2\int_{a_\star}^a\dd a\frac{F(a/\as)}{a^2}
\label{eq:U}
\ee
 where $\mU_\star$ is the initial value of $\mU$. Notice that $\mU$ is the dark electromagnetic energy per unit mass of a given shell characterised by the scale factor $a$. We can rewrite the energy equation \eqref{Eq:Energy1} in a more suggestive form as follows:
 \be
 \frac{\dot{a}^2}{a^2}=\frac{8\pi G\rho_\star}{3 a^3}+2\frac{E(r)-\mU(a)}{a^2}\,.
 \ee
We can then define the inhomogeneous Hubble factor as
\be
H^2(t,r)\equiv\frac{\dot{a}^2}{a^2}=\frac{8\pi G\rho_\star}{3 a^3}+2\frac{E(r)-\mU(a)}{a^2}
\label{Eq:Friedmann1}
\ee
that reproduces the analogous Friedmann equation in an LTB model with appropriate identifications and under some assumptions that we will make explicit in Section \ref{Sec:Lemaitre}. The first term in \eqref{Eq:Friedmann1} is obviously the matter contribution. The energy function $E(r)$ reproduces the contribution from the inhomogeneous spatial curvature $k(r)$. The appropriate interpretation of the potential $\mU$ depends on the particular evolution because it can depend on $t$ and $r$ through its dependence on $a$. In the general case, this potential contributes like an additional component with an effective equation of state determined by \eqref{eq:U}. Since the zero-point of this potential is free, i.e., it will be determined by boundary conditions, there will always be a piece of $\mU$ contributing to the spatial curvature $k(r)$. Before computing the explicit form of these contributions, it is instructive to see what happens in the absence of the screening scale, so $F$ is a constant function (whose value can be absorbed into $\beta$). In that case, we recover that the equation $\eqref{eq:app}$ does not depend on the radial Lagrangian coordinate so the shells co-move, i.e., $a=a(t)$. In that case using \ref{eq:U} we find
\be
\mU=\frac{4\pi G\rho_\star}{3}\beta^2 a^{-1}-\mU_0\,.
\ee
We have obtained the expected result that the Newton's constant is re-dressed by a factor $1-\beta^2$  and the arbitrary zero-point of the potential contributes to the spatial curvature. Crucially, notice that all the $r$-dependence drops and we recover the usual FLRW homogeneous cosmology. 

In the general case, we can compute a reasonable approximation to \eqref{eq:U} without specifying $F$. By setting $\mU_\star=0$ in accordance with our initial conditions\footnote{This may introduce a fine-tuning problem similar to the usual curvature problem of cosmology. Of course, in practice what we are really assuming is that $\mU_\star$ is sufficiently small so that it does not play any role.}, we can introduce $x\equiv a/\as$ and write 
\be
\mU=-\frac{4\pi G\rho_\star}{3\as}\beta^2\int_{x_\star}^x\dd x\frac{F(x)}{x^2} =-\frac{4\pi G\rho_\star}{3\as}\beta^2\int_0^x\dd x\frac{F(x)}{x^2}
\label{splitUexact}
\ee
where we have used that $x_\star=a_\star/\as\ll 1$ to replace the lower limit by 0. In practice, this amounts to removing a finite part that contributes to $U_\star$, but such a contribution is sufficiently  small  to  have no effect. Initially, when the electric force is screened, i.e., for $x<1$ we can assume that the interpolating function takes the form\footnote{Notice that the existence of screening requires $m>0$, while avoiding a divergent potential $\mU$ near the origin requires $m>1$. Physically, this is the necessary condition to avoid a divergent electrostatic energy for point-like sources. Born-Infeld corresponds to $m=2$.} $F(x)\simeq x^m$ so we have
\bea
\nonumber
\mU_{\rm screened} &\simeq&-\frac{4\pi G\rho_\star}{3\as}\beta^2\int_0^x\dd x \, x^{m-2}\\
&=& -\frac{4\pi G\rho_\star}{3\as}\frac{x^{m-1}}{m-1}=-\frac{4\pi G\rho_\star}{3a}\frac{x^m}{m-1}\,.
\label{splitUscreened}
\eea
This expression shows  that the electrostatic energy generates a spatial curvature despite being screened. Furthermore, this contribution grows as $\mU_{\rm screened}\propto a^{m-1}$ and, even though the evolution is self-similar, the spatial curvature already acquires an inhomogeneous profile $\mU_{\rm screened}\propto \as^m\propto r^{m/2}$. In any case, since in the screened region we have $x\ll1$, $\mU$ contributes negligibly with respect to the dust component to the Friedmann equation \eqref{Eq:Friedmann1}. Hence at early times before the shells exit their screening radius, i.e. for $a\ll \as$, the Friedmann equation reduces to the one of the $\Lambda$-CDM model when the spatial curvature vanishes $E\equiv 0$. In this case the scale factor is not affected and grows as $a\propto t^{2/3}$. Only when the screening ceases around $x\simeq 1$ does  this contribution become relevant. In the asymptotic region with $x\gtrsim 1$, the integral can be computed as
\be
\mU\simeq-\frac{4\pi G\rho_\star}{3\as}\beta^2\left[\int_0^1\dd xx^{m-2}+\int_1^x\dd x\, x^{-2}\right]
\label{split}\,.
\ee
The integration can be performed straightforwardly and we obtain
\be
\mU(a)\simeq-\frac{4\pi G\rho_\star}{3\as}\beta^2\left[\frac{m}{m-1}-\frac{\as}{a}\right]
\ee
where we have restored the explicit dependence on $a$. At very late times where $a\gg\as$, the second term becomes negligible and only the first piece contributes. To have a better physical understanding of the two terms, we can insert the expression for $\mU$ into the Friedmann equation \eqref{Eq:Friedmann1} to obtain
\be
H^2(t,r)=\left(1-\beta^2\right)
\frac{8\pi G\rho_\star}{3 a^3}+\frac{m}{m-1}\frac{8\pi G\rho_\star}{3 \as}\beta^2a^{-2}\,.
\label{Eq:Friedmann2}
\ee
We then see that the unscreened region corrects the effective Newton's constant by the factor $1-\beta^2$, as expected because in that region we have the extra electric repulsion, while the screened region contributes a spatial curvature term. This equation can be written as
\be
H^2(t,r)=
\frac{8\pi G\rho_\star}{3 a^3}\left [ 1 +\beta^2 \left(\frac{m}{m-1} \frac{a}{\as}-1\right) \right]\,.
\label{Eq:Friedmann3}
\ee
When the shells have exited their screening radius for $a\ge\as$, the term in $\beta^2$ is always positive. As a result, the correction to the Friedmann equation due to the dark interaction always enhances the Hubble rate compared to the $\Lambda$-CDM case.

It is interesting to notice that the asymptotic evolution is dominated by the growth of $\mU_{\rm screened}$ during the screened phase that saturates at $x\simeq1$ and give an inhomogeneous spatial curvature. Although it may seem like this inhomogeneous spatial curvature induces a scale-dependent expansion rate at late times, this is not the case and it actually gives rise to an asymptotically homogeneous Hubble expansion rate. This can be understood directly from the equation \eqref{eq:app} by noticing that at late times the dominant solution is $a\propto t$ as it corresponds to a curvature dominated Universe. Since the inhomogeneity appears in the proportionality constant, the Hubble expansion rate is not  sensitive to it and  all the shells  enter again a comoving motion, although  an inhomogeneous density profile  has been generated. To see why the curvature contribution from $\mU$ to the Friedmann equation is homogeneous, we can notice that $\as\propto\sqrt{r}$. On the other hand, we can compute the effect of the asymptotic dominance of the curvature term in the Friedmann equation \ref{Eq:Friedmann2} on the expansion rate. In particular we find that asymptotically the scale factor $a_\infty$ acquires a dependence on the radius $r$ as
\bea
\nonumber
a_\infty&=&\frac{a_\infty}{\as}\frac{\as}{a_\star}a_\star=\left(\frac{t_\infty}{t_{\rm s}}\right)\left(\frac{t_{\rm s}}{t_\star}\right)^{2/3}a_\star\\
&\propto & t_{\rm s}^{-1/3}\propto \as^{-1/2}\propto r^{-1/4}
\eea
where we have used that $a\propto t^{2/3}$ in the unscreened phase as the effects of the electrostatic force is null in this era and the Friedmann equation reduces to the one of the $\Lambda$-CDM model when we take a vanishing curvarture $E\equiv 0$ and the scale factor grows in $t$ in the curvature dominated regime. Since the inhomogeneous curvature contribution from $\mU$ in the asymptotic region goes like $a_\infty^{-2}\as^{-1}$ we see that the $r$-scaling exactly cancels. Yet another way of seeing the scale-independence is to notice that the asymptotic Friedmann equation gives $\dot{a}\propto \as^{-1/2}(r)\Rightarrow a\propto \as^{-1/2}(r)t$ so $H=\dot{a}/{a}$ does not depend on $r$. In particular, this means that the asymptotic scaling of the spatial curvature decays as $k_\infty(r)=2\mU_\infty\propto r^{-1/2}$ so the larger scales are less influenced by the produced spatial curvature. Let us emphasise that despite recovering the comoving evolution in the asymptotic late-time region, the cosmological principle is broken due to the inhomogeneous profile for the scale factor of the different scales generated by the scale-dependence of $\as$. This breaking of homogeneity gives observable effects like e.g. on the redshifts of photons as we will discuss in Section \ref{Sec:H0}.

Thus, the overall evolution exhibits three phases:
\begin{itemize}
    \item {\it Phase 1: Comoving dust dominated evolution.} The first stage of the evolution is insensitive to the electric force, which is screened on all scales, and the Universe evolves in comoving motion. However, the inhomogeneous contribution from $\mU$ to the spatial curvature already grows
    \item {\it Phase 2: Transition region.} Some scales start exiting their screening radii so they decouple from the comoving motion due to the additional electric force. Since the screening scale factor scales as $\propto \sqrt{r}$, smaller scales decouple from the comoving motion at earlier times.
    \item {\it Phase 3: Asymptotic comoving inhomogeneous evolution.} At very late times, when all the relevant scales have exited their screening radii, the comoving motion is recovered, but for an inhomogeneous density profile formed during phase 2. This asymptotic state is dominated by the spatial curvature associated to the electric potential that has been growing since phase 1.
\end{itemize}
We will confirm these findings in the numerical solutions of Section \ref{Sec:numerics}. In the next section we will see how our Newtonian picture relates to relativistic inhomogeneous cosmological models. But before that and for completeness, let us give the corresponding expressions when the mass is not conserved, i.e., for $\mu=\mu(t,r)$. In that case we can still write a first integral of \eqref{Eq:appgeneral} as
\be
E(r)=\frac12 \dot{a}^2-\frac{G\mM(a)}{a}+\tilde{\mU}(a)
\ee
where the functions $\mM$ and $\tilde{\mU}$ satisfy
\be
\mM-a\frac{d\mM}{d a}=\mu(a),\quad\text{and}\quad \frac{d \mU}{d a}=-\beta^2\frac{G\mu(a)}{a^2} F(a/\as)
\ee
and $\mu$ must be interpreted as a function of $a$. The function $\mathcal{M}$ will be identified with the mass in the shell labelled by $a$ and $\tilde{\mathcal{U}}$ is related to the electrostatic potential which will be shown to correspond too the pressure due to the electrostatic interaction in the next section. It is not difficult to see that for $\mu=\mu_\star$ we recover our previous results when mass conservation holds.

\subsection{Connection to Lema\^itre models}
\label{Sec:Lemaitre}

In this section we will show how, with suitable identifications, the dynamics of  charged DM admits a geometrical interpretation in terms of an inhomogeneous, spherically symmetric metric. To do so, we will introduce a dark fluid as a proxy for the description of the properties of non-linear electrodynamics cosmologically. This dark fluid will not have an equation of state equal to one-third as in linear electrodynamics. At the field theory level, this follows from the non-vanishing of the trace of the energy-momentum tensor for non-linear electromagnetism that measures the breaking of scale invariance. Moreover we will see that the pressure must also be inhomogeneous. This can be described using Lema\^itre models    \cite{Lemaitre:1933gd}. A fully relativistic description where the origin of the dark fluid will be shown to follow from the presence of an isotropic electric field is given in \cite{BeltranJimenez:2021imo}. 

\subsubsection{The Einstein equations}

The equations derived in the previous section from the Newtonian approach can be matched, with appropriate identifications, to those derived from a Lema\^itre model \cite{Lemaitre:1933gd}. This belongs to the category of inhomogeneous spherically symmetric solution to the Einstein equations (see \cite{Bolejko:2011jc} for a review) which has been extensively applied to cosmology \cite{Enqvist:2007vb}. The Lema\^itre metric reduces to the well known Lema\^itre-Tolman-Bondi (LTB) one \cite{Lemaitre:1933gd,Tolman:1939jz,Bondi:1947fta} if one takes a Universe filled only with dust (zero pressure) and a cosmological constant term. Generalisation of LTB models to include a time \textit{or} space dependent pressure have been investigated as well \cite{Lasky:2010vn,Grande:2011hm,Lynden-Bell:2016nys}.
However, in order to treat with full generality a time \textit{and} space dependent pressure, one needs to resort to the Lema\^itre model.

The metric for this class of models is given by the following line element
\begin{equation}\label{eq:metric_Lem}
    \dd s^2 = - e^{A(r,t)}\dd t^2 + e^{B(r,t)}\dd r^2 + R^2(r,t)\dd\Omega^2
\end{equation}
where $d\Omega$ is the solid angle that can be fixed once an origin for the coordinates has been chosen.

We will assume that the matter content can be described as a perfect fluid and  comprises a pressure-less dust, a (dark)  fluid with pressure and a cosmological constant term. We will further consider non-interacting fields, hence their individual stress-energy tensor are covariantly conserved.

The Einstein equations for the metric \eqref{eq:metric_Lem} read
\begin{eqnarray}
\label{eq:B}
2\frac{\dot R'}{R}-\frac{\dot B R'}{R}-\frac{A'\dot R}{R} & =& 0\,,\\
\label{eq:EFE1}
 4\pi R^2\dot R p_{\rm tot}(r,t)& = & - \dot M_{\rm tot}\,,\\
 \label{eq:EFE2}
 4\pi R^2 R' \rho_{\rm tot}(r,t) &=&  M_{\rm tot}'
\end{eqnarray}
where
\begin{equation}\label{eq:mass_def}
    2 G M_{\rm tot} = R + R e^{-A} \dot R^2 - R e^{-B} R'^2 - \frac{1}{3}\Lambda R^3
\end{equation}
is the Lema\^itre \cite{Lemaitre:1933gd} or Misner-Sharp-Hernandez mass \cite{Misner:1964je,Hernandez:1966zia}. Notice that $\Lambda$ has dimension two in natural units where $\hbar =1, \ c=1$. Hence  
here $\rho_{\rm tot}$ and $p_{\rm tot}$ denote the total energy density  and pressure of all the fluids in the Universe but the one representing the cosmological constant. 

\subsubsection{Conservation of matter}

From the conservation of the energy momentum tensor we get
\begin{equation}\label{eq:cont}
    \dot B + 4 \frac{\dot R}{R} = -\frac{2\dot \rho_{\rm tot}}{\rho_{\rm tot}+p_{\rm tot}} \qquad {\rm and} \qquad A' = -\frac{2 p_{\rm tot}'}{\rho_{\rm tot} + p_{\rm tot}}\,.
\end{equation}
The second equation clearly shows how the gradients of the pressure source the function  $A$ and fully determines it. More particularly, we have
\begin{equation}
    A=-\int \dd r \frac{2 p_{\rm tot}'}{\rho_{\rm tot} + p_{\rm tot}} + \tilde A(t)
\end{equation}
where $\tilde A (t)$ depends only on time and is not determined. In fact, we can always choose $\tilde A\equiv 0$ as this can be absorbed in a change of time $t \to \tilde t$ where $\dd\tilde t= \dd t e^{\tilde A/2}$. Also notice  that, 
in the absence of pressure term we retrieve the fact that $A\equiv 0$ like in Lema\^itre-Tolman-Bondi models. 
Joining equations \eqref{eq:EFE1} and \eqref{eq:EFE2} with those in \eqref{eq:cont} and recalling that the two fluid species are not interacting, i.e. their stress energy tensor are separately conserved in an inhomogeneous Universe, we get the continuity equations in this class of models to be
\bea
 \dot \rho_{\rm dust}&+& \left(2\frac{\dot R}{R} + \frac{R'}{R}\right) \rho_{\rm dust}=0\,,
\\
 \dot \rho_{\rm e}&+& \left(2\frac{\dot R}{R} + \frac{R'}{R}\right) (\rho_{\rm e}+ p_{\rm e})=- \frac{\dot R}{R'} p'_{\rm e}\,.
\label{eq:cont_split}
\eea
where we have explicitly separated the total energy density $\rho_{\rm tot}=\rho_{\rm dust}+\rho_{\rm e}$ where $\rho_{\rm dust}$ corresponds to the pressureless fluid contribution while $\rho_{\rm e}$ is the energy density of a fluid with an inhomogeneous pressure whose origin will be identified as representing the effects of the electrostatic interaction, i.e. the dark fluid.
The first equation in (\ref{eq:cont_split}) expresses the conservation of the dust component of the matter content. The second equation follows from the conservation of dust in an inhomogeneous Universe and \eqref{eq:EFE1} together with \eqref{eq:EFE2}. Finally, notice that the above system of equations reduce to the standard ones for the homogeneous case if $p'=0$ and $R=r a(t)$. 

Equation \eqref{eq:B} can be integrated and gives
\bea
\nonumber
   e^{B(r,t)} &=& \frac{R'^2}{1 +2 E(r)}e^{-\int A'\frac{\dot R}{R'}\dd t}\\
   &=& \frac{R'^2}{1 +2 E(r)}e^{2\int \frac{p_{\rm tot}'}{\rho_{\rm tot}+p_{\rm tot}}\frac{\dot R}{R'}\dd t}
   \label{Beqw}
\eea
where we have used the second equation in \eqref{eq:cont}. 

\subsubsection{The Friedmann equation}

The mass defined in equation \eqref{eq:mass_def} can be conveniently rewritten as
\begin{equation}\label{eq:massRdot}
    e^{-A}\dot R^2= \frac{2G  M_{\rm tot}}{R}+\frac{1}{3}\Lambda R^2 +  (1+ 2 E(r))e^{-2I_B}-1
\end{equation}
where we have introduced the integral
\begin{equation}
    I_B=\int\frac{p_{\rm tot}'}{ \rho_{\rm tot}+p_{\rm tot}}\frac{\dot R}{R'}\dd t\,.
\end{equation}

The total mass appearing in the above equation can be separated in its two components
\bea
\nonumber
     M_{\rm tot} (r,t) &=& 4\pi\int \Big[\rho_{\rm dust}(r,t)+\rho_{\rm e}(r,t)\Big] R^2(r,t) R'(r,t)\dd r\\
     &=& M_N(r,t) + M_{\rm e}(r,t)\,,
\eea
where in the last equality we have defined the Newtonian mass $M_N$ associated to the dust component and the electrostatic mass $M_{\rm e}$.
Notice that the time derivative of $M_N$ vanishes thanks to the conservation of $\rho_{\rm dust}$. If we also assume that there is no shell crossing, i.e.  $R'(r,t)\neq 0$, then $M_N$ is conserved inside the sphere of radius $R(r,t)$ at any given time. It is then enough to provide its value at some reference time. This is not true for the second mass $M_{\rm e}$, since the presence of the pressure implies the non-conservation of the mass. 

Let us now consider the situation in which $p_{\rm e}\ll \rho_{\rm tot}$ and the the dimensionless ratio of  pressure gradients over the total energy density is not too large compared to the inverse size  $R^{-1}$, i.e. we consider that this term varies slowly over the whole shell.
As both $A(r,t)$ and $B(r,t)$ depend on  integrals of the quantity $p_{\rm e}'/(\rho_{\rm tot} +p_{\rm e})$,  we can expand the exponential and consider only the leading terms. This gives
\begin{equation}\label{eq:lam_Newt}
  (1-A)\frac{\dot R^2}{R^2}= \frac{2 G  M_{\rm tot}}{R^3}+\frac{1}{3}\Lambda + 2\frac{ E(r)- I_B}{R^2}
\end{equation}
where $\Lambda=8\pi G  \rho_{\rm DE}$ is related to the dark energy scale $\rho_{\rm DE}$. 
One can see that equations \eqref{eq:lam_Newt} is formally equivalent with the one obtained from the Newtonian cosmology approach. Indeed let us identify 
\be 
\frac{I_B}{R^2}=\frac{\mU(a)}{a^2}
\ee
where $R(r,t)=a(r,t) r$ for each shell and $\mU(a)$ is given by \eqref{eq:U}. This gives
\be 
\frac{p_{\rm e}'}{\rho_{\rm tot}+p_{\rm e}}=\frac{4\pi \beta^2 G  \rho_* rR'}{3a^2(r,t)}F\left(\frac{a}{a_s}\right)
\label{eqfirst}\,.
\ee
The identification with the Newtonian cosmology approach is valid when the dust component dominates over the dark fluid and we have $\rho_{\rm tot}+p_{\rm e}\simeq \rho_{\rm dust}$. We can see that (\ref{eqfirst}) then determines $p_{\rm e}$
which is proportional to  ${\cal O}(G )$ from (\ref{eqfirst}) implying that the term in $2G  M_{\rm e}/R^3$ in the Friedmann becomes, thanks to (\ref{eq:EFE2}), of order $G ^2$.
This is a post-Newtonian term as the Newtonian approximation can be seen as an expansion in powers of $G$ and small velocities. Hence this term  goes beyond the Newtonian approximation, being second order,  and can be dropped in the identification with the Newtonian cosmology equations. Thus we can discard  the term in $M_e$ in the total mass as well as the contribution due to the potential $A$ and keep the Friedmann equation at the lowest order
\be
\frac{\dot R^2}{R^2}= \frac{2 G  M_{N}}{R^3}+\frac{1}{3}\Lambda + 2\frac{ E(r)- I_B}{R^2}
\ee
which completes the identification between the Newtonian approximation of the Lema\^itre model and the cosmological model with a dark electromagnetic component presented in the previous section. In agreement with our previous definition, we have identified the mass $M_N$ as the mass inside the shell of radius $R$ due entirely to the dust component of the Universe. This is exactly  the mass term that appears in the Newtonian derivation of the Friedmann equation where a shell of mass $R$ evolves under the influence of the gravitational potential inside its radius. This comes eventually from the effacement theorem which is true  for Newtonian gravity and can be extended to screened electrostatic interaction, see appendix \ref{app:A}. This equivalence is only valid when no shell crossing happens and only one type of charged matter is present. A good example is provided by the Born-Infeld theory.  More complex cases with charged and uncharged species, together with shell crossing will be considered below.

\subsubsection{Thermodynamic interpretation}

The Einstein equation \eqref{eq:EFE2} can be used to identify the mass of the pressure component as
\be 
M_{\rm e}(r,t)= -4\pi \int \dd t\, p_{\rm e} R^2 \dot R\,.
\label{secondM}
\ee 
Notice that the dark radiation mass is equal to the work due to the dark pressure. 
In this sense we can identify the pressure as
\be 
p_{\rm e}= -\frac{\dd M_{\rm tot}}{\dd V}
\ee
where the volume of a given shell labelled by $r$  is such that $\dd V= 4\pi R^2 \dot R \dd t= 4\pi R^2 \dd R$.
This can also be written as the pressure due to the electrostatic force on a given shell as
\be 
p_{\rm e}= \frac{F_{\rm e}}{S}
\ee
where the electrostatic force on a  shell
\be 
F_{\rm e}= -\frac{\dd U_{\rm e}}{\dd R}
\ee
is the gradient of the internal energy $U_{\rm e}$ identified with
\be
U_{\rm e}\equiv M_{\rm tot}
\ee
and $S=4\pi R^2$ is the surface area of the given shell at time $t$. Hence, we see that the dark radiation pressure is responsible for the variation of the internal energy, i.e. the total mass, of a given shell. In the Newtonian approximation, the electrostatic pressure can be easily computed as
the force acting on a given shell of radius $R$ divided by its surface area $4\pi R^2$. The electrostatic force exerted on the shell of radius is given by 
\begin{equation}
    F_{\rm e}= \frac{\beta^2 G M^2(R)}{ R^2} F(R/\rs)
\end{equation}
from (\ref{eq:dyn})
leading to the pressure $p_{\rm e}=F_{\rm e}/4\pi R^2$
\be
p_{\rm e }=\frac{\beta^2 G M^2(R)}{4\pi R^4} F(R/\rs)=\beta^2\frac{4\pi G\rho_\star^2}{9a^4}r^2F(a/\as)\,.
\label{Eq:peNewtonian}
\ee
We can compare this expression to the pressure in the relativistic version. If we consider \eqref{eqfirst} and take a Newtonian limit so we can neglect $p_{\rm e}$ against $\rho_{\rm tot}=\rho_\star/a^3$ and use that $R'=a$ we obtain:
\be
p'_{\rm e}=\frac{4\pi \beta^2 G\rho_\star^2}{3a^4}rF(a/\as)\,.
\ee
In the asymptotic region, we can now recall that $a\propto r^{-1/4}$ and approximate $F\simeq 1$ so we can neglect its $r$-dependence. Under these assumptions, we can integrate the above expression and explicitly check that we obtain the pressure \eqref{Eq:peNewtonian} already known  from the Newtonian approach. Notice that the dark interaction contributes to the Friedmann equation because the pressure of the dark fluid has a non-vanishing gradient. This gives a contribution to the curvature of space. Hence we see explicitly that the effects of the dark force goes beyond the usual treatment of inhomogeneities using Lema\^itre-Tolman-Bondi space-times, and necessitates the more general Lema\^itre models.  

\section{Numerical results}
\label{Sec:numerics}

In this Section we will solve numerically the evolution of the shells and we will confirm the analytical findings of the precedent section. We will proceed in several steps to clearly identify the different effects. To that end, we will first introduce our numerical approach to the problem and solve it for a single component scenario as we did in the previous analytical analysis. The numerical solutions will allow us to  make  explicit when the previous assumption of the absence of shell crossing, even though the additional electric force is repulsive, is valid. We will then proceed to include uncharged baryons and dark energy and discuss how this scenario can be relevant to alleviate the $H_0$ tension.

\subsection{Evolution of Dark Matter halos}

We can confirm the phenomenology explained above from the Newtonian cosmology approach by  solving numerically the evolution of many shells. We will tackle the numerical problem by discretising the shells distribution so we will consider a set of $N$ shells with radii $R_i(t)$ subject to the following system of equations
\be
\label{eq:anim}
\ddot{R}_i=-\frac{GM(R_i)}{R_i^2}\Big[1-\beta^2 F(R_i/r_{{\rm s},i})\Big],\quad\quad i=1,\cdots,N.
\ee
We will solve these equations for a set of discrete shells with a uniform initial distribution, i.e., assuming a homogeneous density profile initially.
In all the cases that we consider, we have checked that for large values of $N$, typically one hundred, the evolution of the shells converges to an asymptotic behaviour which corresponds to the analytical understanding that we have just presented. In particular, in the absence of shell-crossing we expect the large $N$ limit to converge to the description we have given in terms of Newtonian cosmology. Because the spherical symmetry is preserved in the evolution, the mass for each shell $m_i$ will remain constant. Notice that this does not mean that the function $M(R_i)$, which is the mass contained within the $i$-th shell, is constant in the evolution. This only happens in the absence of shell-crossing. Numerically, we compute the mass $M(R_i)$ as
\be
M(R_i)=\sum_{R_j\leq R_i}m_j
\ee
where the time-dependence comes from the time dependence of the summation limits because, as explained, $m_i$ are constant. Obviously, if there is no shell-crossing, the summation limits do not depend on time and the mass enclosed by each shell is conserved. If the initial uniform density is $\rho_\star$, we assign to each shell the mass contained between that shell and the immediate inner one in the initial distribution so that
\be
m_i=\frac{4\pi\rho_\star}{3}\Big(R_i^3-R_{i-1}^3\Big)\Big\vert_{t=\tini},\quad\quad i=1,\cdots,N
\ee
with $R_0=0$. Because of the initial uniform density profile, this mass assignment guarantees that $M(R_i(\tini))\propto R_i^3(\tini)$ which establishes an initial hierarchy for the screening radii. Finally, for the initial velocities we will assume that the shells are in a comoving regime with $v_i(\tini)=H_\star R_i(\tini)$. These initial conditions are motivated as at early times we impose that all the shells are inside their screening radii so that it is natural that they evolve according to the usual Hubble flow. We also impose the spatial curvature contribution to be negligible with respect to the dust energy density so we take $H_\star=\sqrt{\frac{8\pi G\rho_\star}{3}}$. This initial velocity guarantees that we will pick the solution corresponding to a dust dominated expansion.

\begin{figure*}[ht]
\includegraphics[width=0.49\linewidth]{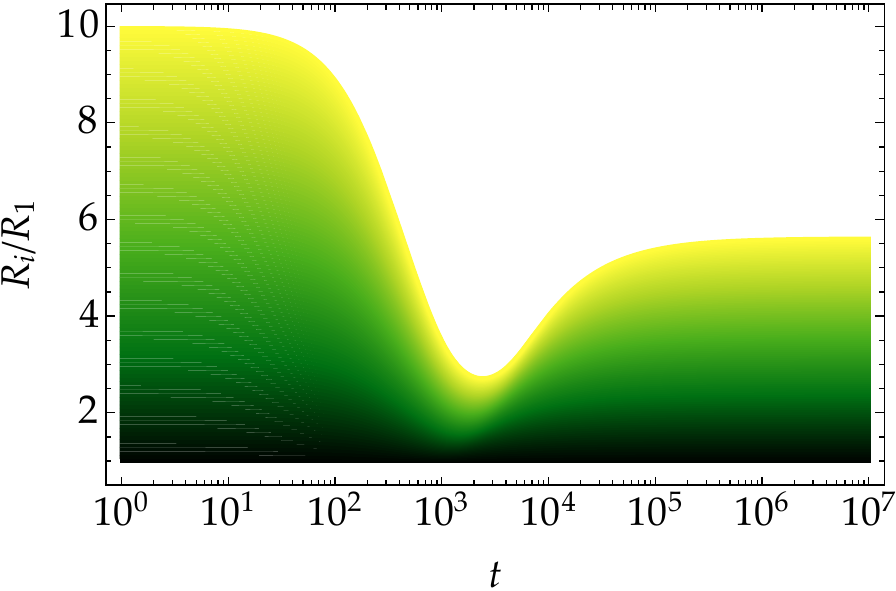}
\includegraphics[width=0.49\linewidth]{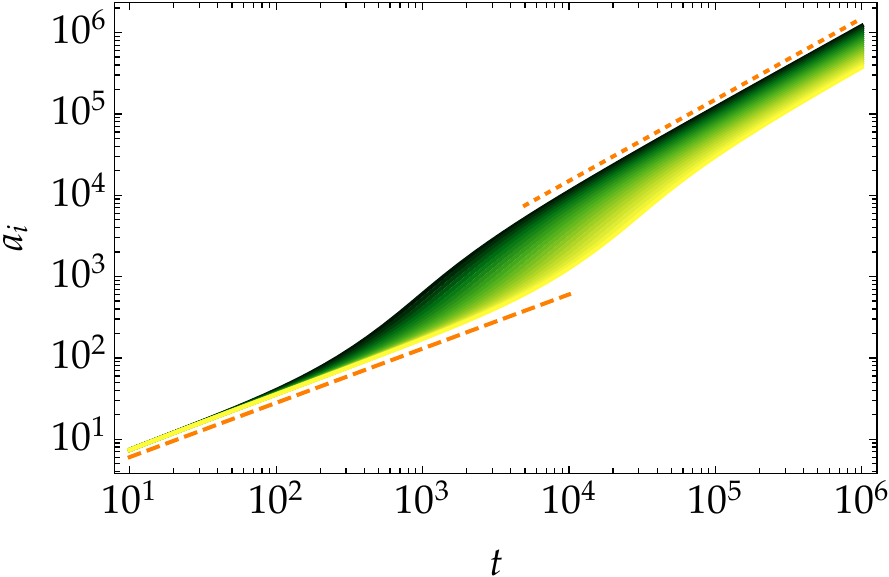}
\includegraphics[width=0.49\linewidth]{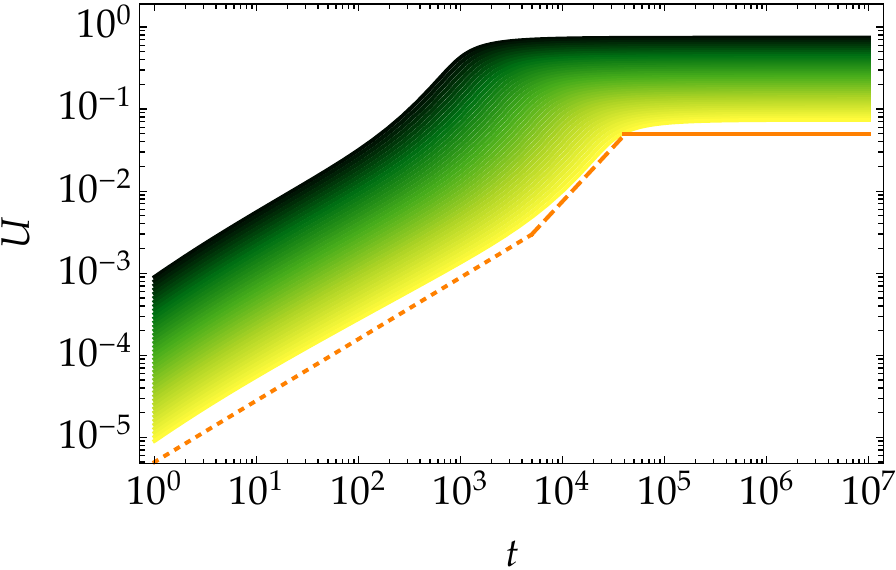}
\includegraphics[width=0.49\linewidth]{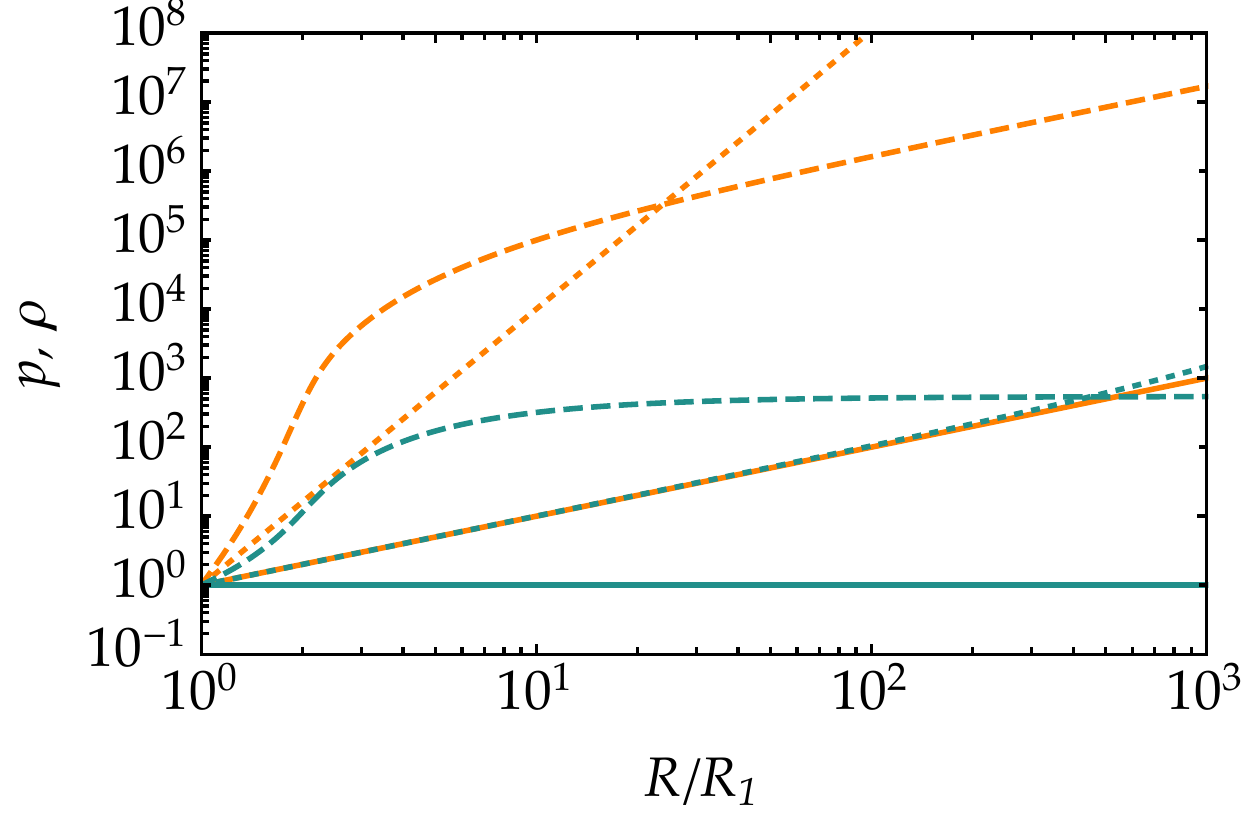}
\caption{The following choice of values are only taken for illustration so that the effects can be visible.  In the upper left panel we show the evolution of the shells size normalised to the radius of the innermost shell. The colour coding has been chosen so that darker means inner shells. We have chosen $\beta^2=200$, $4\pi\rho_\star/3=1$ and $\rs=500\sqrt{M}$ in units of $G=1$. In the upper right panel we show the shells scale factor evolution, where we can see the initial dust dominated evolution (dashed line) that transits to an open Universe (dotted line) with a spatial curvature generated by the electric pressure. We can also confirm the initial self-similar evolution that breaks as the different shells exit their corresponding screening radii. The lower left panel shows the time evolution of the function $\mU$ which governs the evolution of the curvature in the effective Friedmann equation. Finally, in the lower right panel we plot the density (blue) and the pressure (orange) profiles at three different times: the initial profile (solid line), the profile at a time in the transition region, $t=3\times 10^3$, (dashed) and the asymptotic profile (dotted). This confirms that the asymptotic state recovers the self-similar evolution but an inhomogeneous profile has been generated. Moreover we also see explicitly that the pressure acts on the outer shells at late times. }
\label{Fig:riBI}
\end{figure*}

For our numerical solutions, we will use the Born-Infeld model, so the interpolation function is $F(x)=(1+x^{-4})^{-1/2}$ and we show the evolution of the shells in Figure \ref{Fig:riBI}.  In the upper left panel we can see the evolution of the shells size normalised to the (time-dependent) radius of the innermost shell, while the upper right panel shows the evolution of their scale factors. The three phases described above are clearly visible, namely: the shells comove initially with a common scale factor that grows as $a\propto t^{2/3}$, as it corresponds to dust domination. When the shells start exiting their respective screening scales the comoving evolution ceases and different shells start evolving differently. Finally, at sufficiently late times, the shells enter a comoving evolution where they all grow as $a\propto t$, but the breaking of self-similar evolution during the transition phase introduces an inhomogeneous distribution for the scale factors. This inhomogeneity can be identified in the lower right panel where the density profiles at different times are depicted. In particular, we can see a growing density profile in the asymptotic region, in accordance with the fact that the electric force stacks more densely the initial distribution of shells (upper left panel). Finally, we have plotted the potential $\mU$ that gives rise to an inhomogeneous spatial curvature that becomes dominant at late times as it eventually drives the expansion of the shells. As obtained analytically above, there is an initial growth during the screened phase that saturates when the screening ceases (with a transition region). The asymptotic spatial curvature can be seen to be larger for the inner shells, in agreement with the obtained profile $k_\infty\propto r^{-1/2}$.


\subsection{Shell crossing}
\label{Sec:shellcrossing}
As commented above, an important issue in the evolution of the shells is the possibility of having a singular stream flow where different shells cross. This condition in turn determines whether the mass inside a given shell is conserved or not. In this section we will  analyse this issue in more detail with our numerical solutions. Our main purpose here is not to present an exhaustive analysis of the regular evolution of the outwards/inwards matter streams that can give rise to the presence or absence of shell crossing, but rather to show how this crucially depends on the electric force profile with an explicit example. That shell crossing is not always a feature of our scenario should be clear from our analysis in the precedent Section for the Born-Infeld theory. In this case, no shell crossing was present.

The crucial importance of the electric force for having shell crossing essentially lies in both its relative strength, measured by $\beta^2$, and in the interpolation region, i.e., how smoothly or suddenly the transition occurs. While the impact of $\beta^2$ is quite obvious, the dependence of shell crossing on the details of the transition region is less obvious. For a very sudden transition, when a given shell exits its screening radius, the outer shells which are with a slightly larger radius have not exited their screening radius when the inner shells experience an additional repulsion. This may result in the shell crossing phenomenon if the inner shell gets pushed towards the outer shells quick enough, which fully depends on the magnitude of $\beta^2$. On the other hand, if the transition is sufficiently smooth, the effect of the additional repulsive force kicks in more gradually and the precise shape of the force is also relevant. This means that, when the inner shell exits its screening radius, the outer shell already feels a little bit of the repulsion. Hence the relative acceleration between both shells is smaller for smoother transitions. It can happen then that, for a sufficiently smooth transition, the outer shell exits its screening radius before the inner shells has time to catch up with it. This is the general interplay between strength of the interaction $\beta^2$ and smoothness of the transition that determines whether shell-crossing will occur or not.

\begin{figure*}
\includegraphics[width=0.49\linewidth]{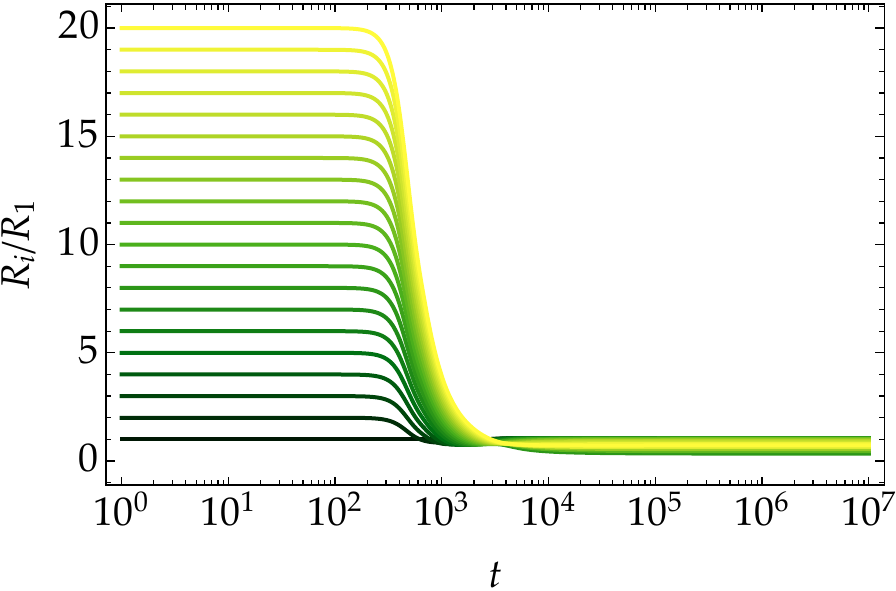}
\includegraphics[width=0.49\linewidth]{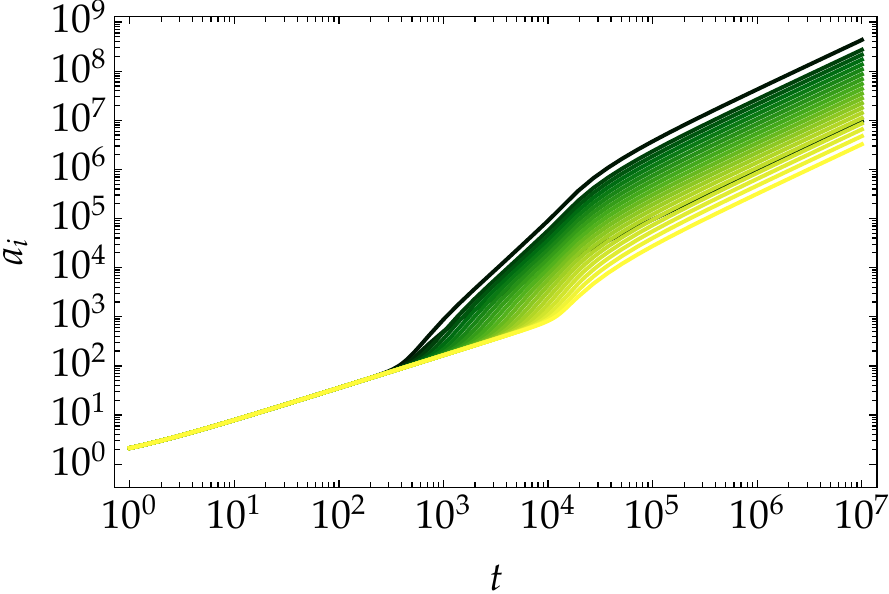}
\includegraphics[width=0.49\linewidth]{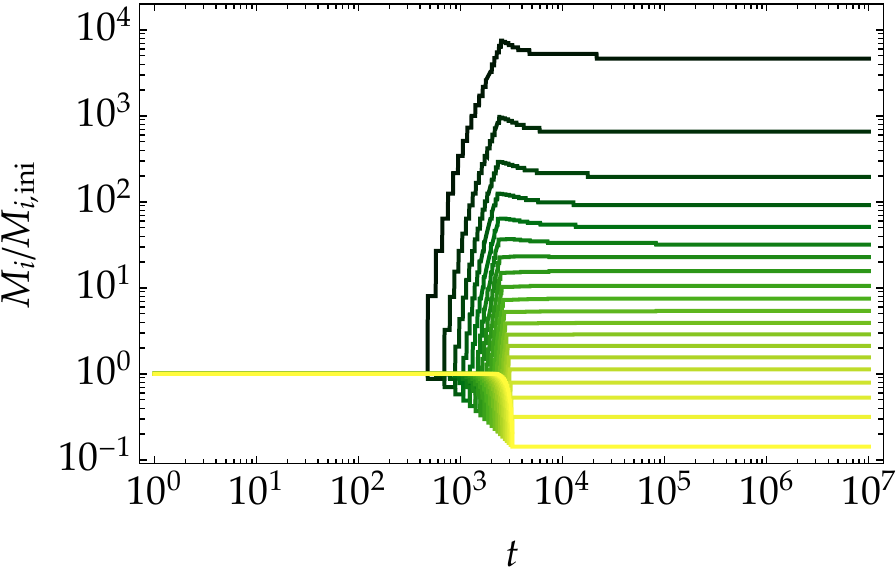}
\includegraphics[width=0.49\linewidth]{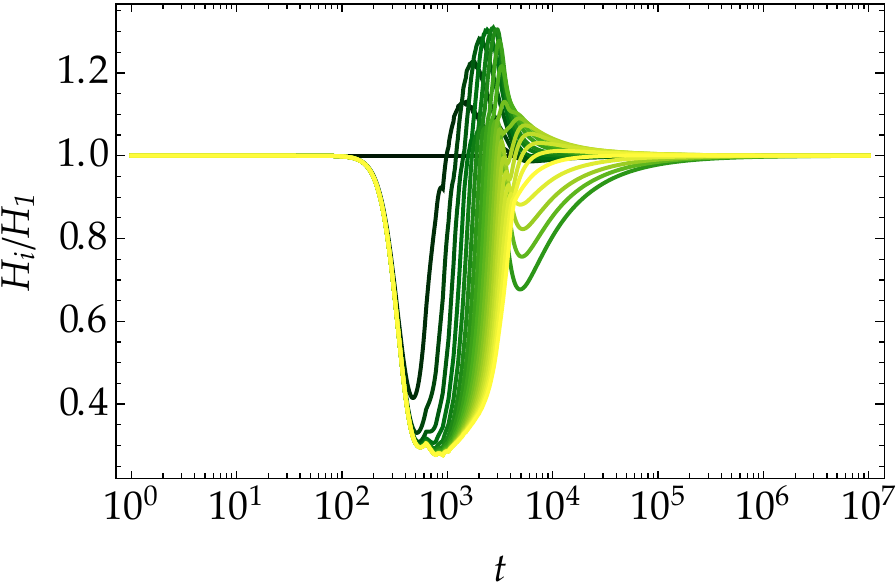}
\caption{In this figure we show a case exhibiting shell crossing. We have chosen the same values as in Fig. \ref{Fig:riBI} except for screening scale that has now been chosen as $\rs=100\sqrt{M}$. The colour coding is also the same. We can see how the initially innermost shell starts gaining mass when it exists its screening radius so it becomes the most massive shell in the asymptotic state, in agreement with the fact that it overtakes all the shells and it becomes the outermost one at late times. On the other hand, the outermost shells lose mass and eventually become the innermost shells.}
\label{Fig:Mi}
\end{figure*}

In order to illustrate the different regimes, we will use 
an artificially modified Born-Infeld electromagnetism parameterisation as a proxy for an electric force whose interpolating function is given by
\be
F=\frac{1}{\left[1+\left(\frac{\rs}{R_i}\right)^4\right]^2}\,.
\ee
This profile is sharper than the pure Born-Infeld one and, consequently, it is more prone to exhibit shell crossing. In Figure \ref{Fig:Mi} we show the evolution and we can clearly see the appearance of shell crossing. It is important to emphasise that this only happens above a certain value for $\beta^2$, while if $\beta$ is sufficiently small, shell crossing can still be prevented. The uniformly distributed shells initially evolve in comoving motion with $a\propto t^{2/3}$ as before. However, when they start exiting their screening scales we can observe how the evolution in the transition region differs from the Born-Infeld case in Figure \ref{Fig:riBI} and now the different shells cross. Eventually, the comoving motion is again recovered, but the generated inhomogeneity is crucially different. In particular, we can see how the shells become substantially more densely distributed, indicating a much steeper asymptotic density profile. 

We also show how the mass function for each shell evolves in time. While for the Born-Infeld model with no shell-crossing the mass is conserved, the model that exhibits shell-crossing leads to an evolution where the mass of each shell is not conserved due to the the gain/loss of mass of the shells as they absorb or exit other shells. The shell crossing as well as the mass variation of the shells can be seen in figure \ref{Fig:shells} top panel and in the animations available at \cite{shells}.

We should warn that the mass evolution shown in Figure $\ref{Fig:Mi}$ shows an effect due to having considered a finite number of shells. This obviously affects the mass profile, but also the saturation of the innermost shell that eventually becomes the outermost one. As we can see, the mass saturates when the shell has overtaken all the shells. By including a wider range of shell sizes, the asymptotic masses of the different shells will change. This will also impact the evolution of those shells since they can keep increasing their mass for a longer time so the transition phase is longer. In any case, let us repeat once more that our aim here is to provide an explicit example of shell-crossing and not to perform an exhaustive analysis so we will not enter into a more detailed analysis of the shell crossing and we will content ourselves with signalling its relevance for our purposes. The interest of exposing the possibility of shell crossing in our scenario is to highlight the crucial difference with the standard case where there is no shell crossing. It is important to emphasise in this respect that we are considering an initially isotropic an homogeneous distribution so that the shell crossing is genuinely produced by the dark electric repulsion. An inhomogeneous density profile or an anisotropic distribution in the initial configuration can also give rise to shell crossing when the shells evolve solely under the influence of gravity.

To finalise our discussion on shell crossing, it is important to notice that the explained casuistic occurs because the additional force is repulsive. Since the screening models based on scalar fields give rise to an additional attractive force, shell crossing does not take place because each shell slows down once it exits its screening radius, thus working in the precise opposite direction, i.e., it actually helps to prevent shell-crossing. The inhomogeneisation of an initial uniform profile can still persist however and this has an impact on the predicted mass function from the spherical collapse as computed with the Press-Schechter formalism.


\subsection{Adding baryons}
\label{Sec:baryons}

\begin{figure*}
\includegraphics[width=0.49\linewidth]{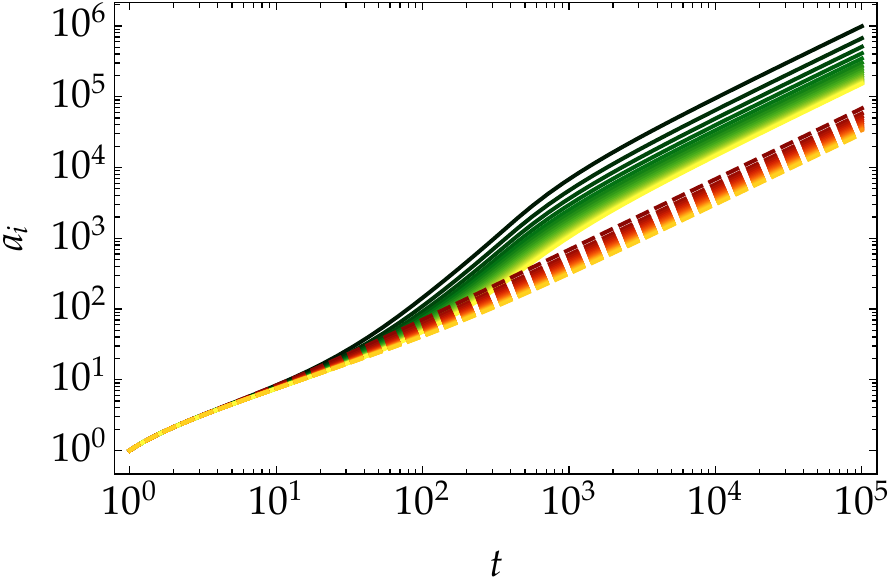}
\includegraphics[width=0.49\linewidth]{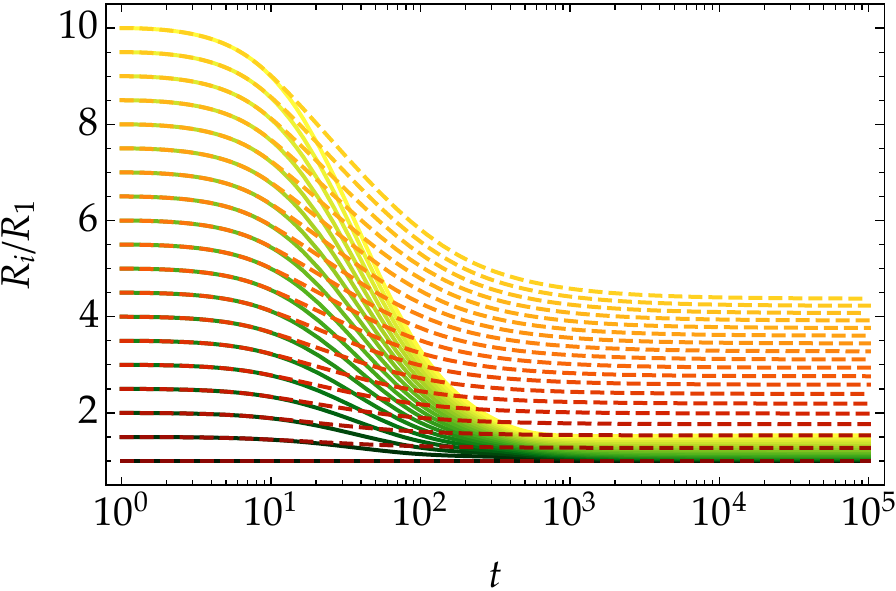}
\includegraphics[width=0.49\linewidth]{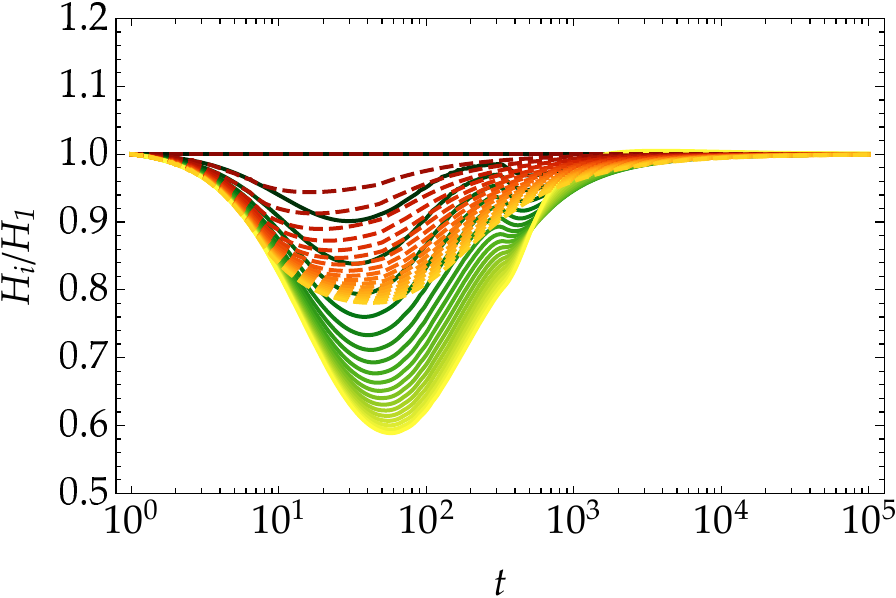}
\includegraphics[width=0.49\linewidth]{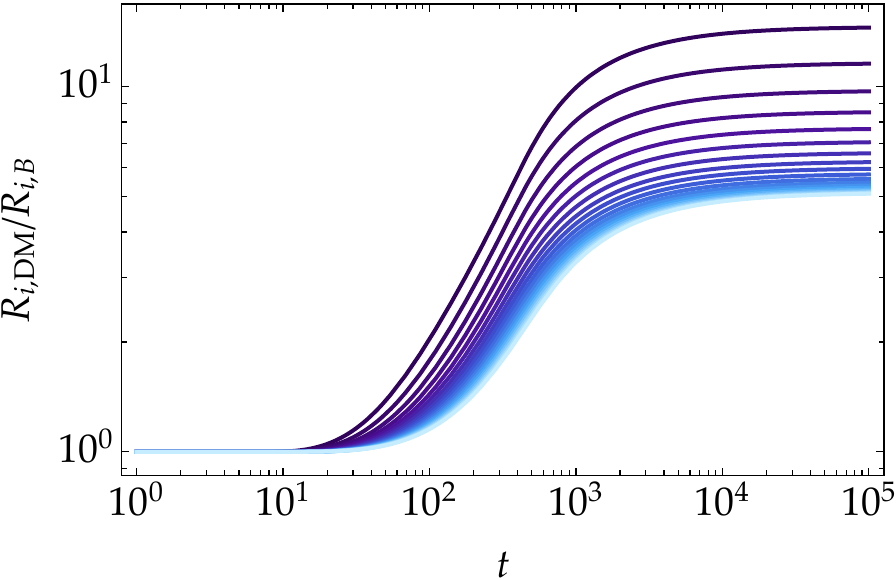}
\caption{The upper left panel shows the scale factor evolution for the DM (solid-green palette) and  baryons (dashed-red palette) shells. In the upper right panel we plot the evolution of the shell's size relative to the size of the innermost shell of each component. The lower left panel shows the expansion rates for both components normalised to the Hubble factor of the innermost shell of each component. Finally, in the lower right panel we can see how the relative size of the initially comoving shells of baryons and DM evolves.}
\label{Fig:RBDM}
\end{figure*}

\begin{figure*}
\includegraphics[width=.4\linewidth]{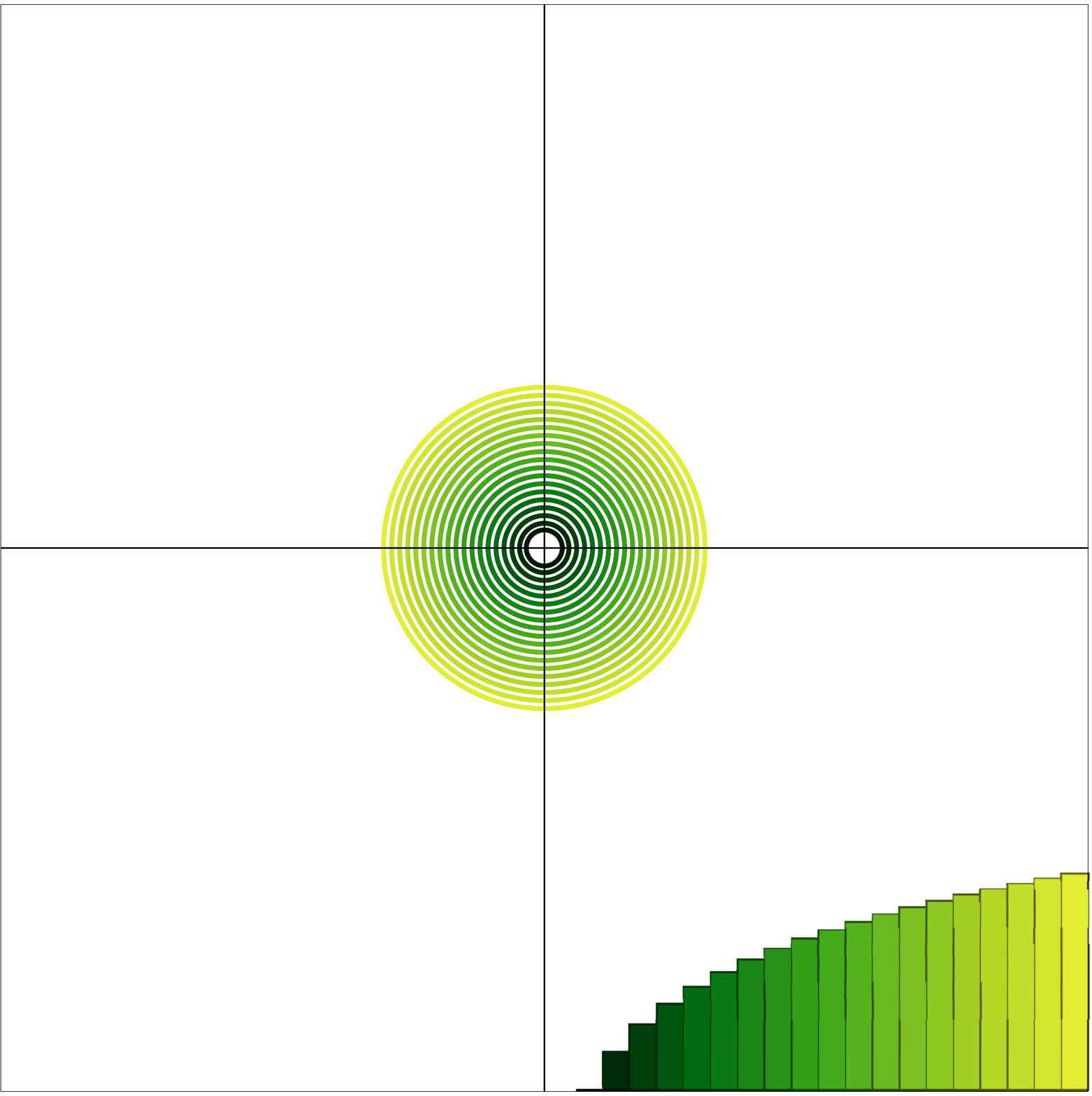}
\includegraphics[width=0.4\linewidth]{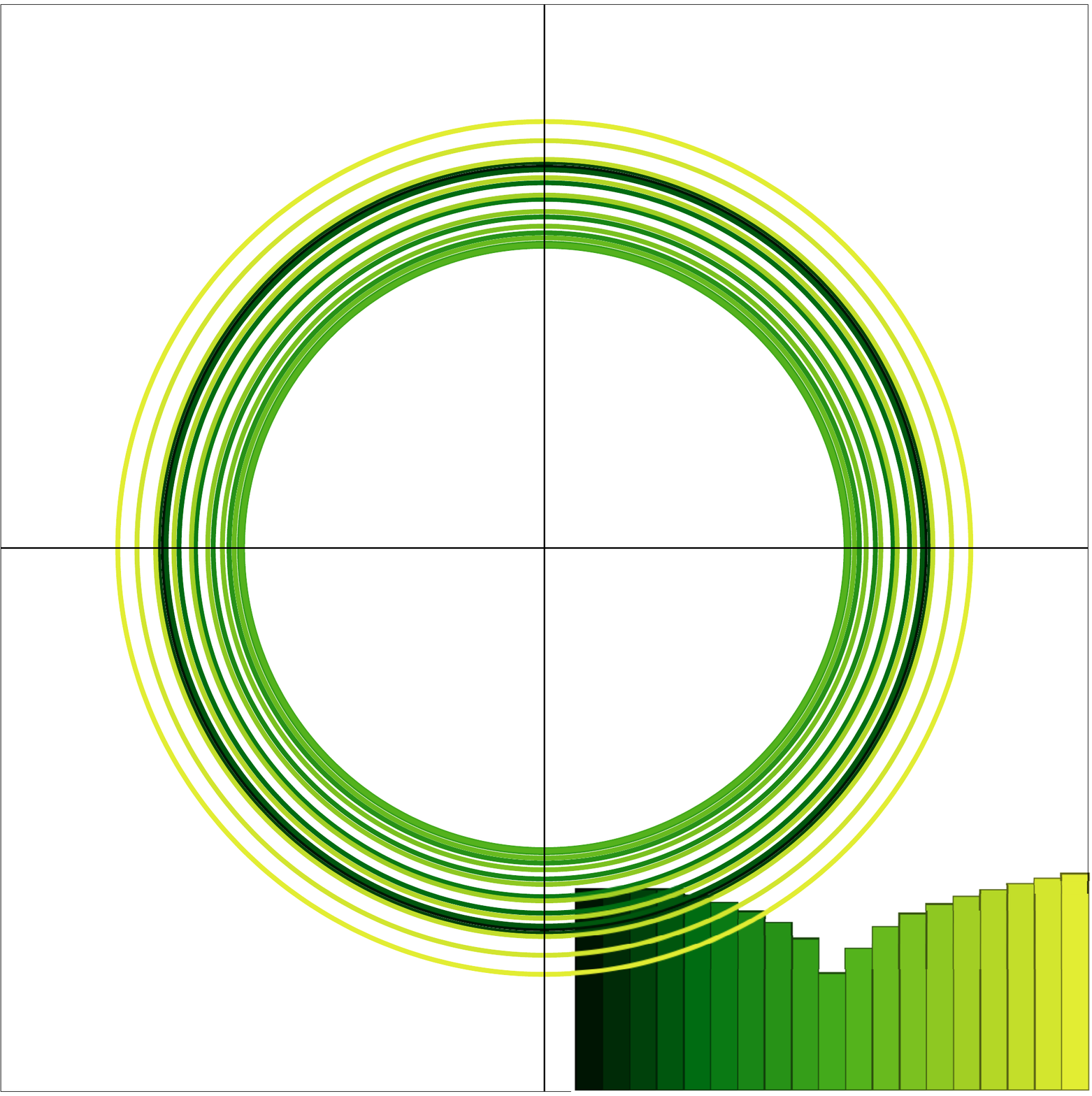}
\includegraphics[width=0.4\linewidth]{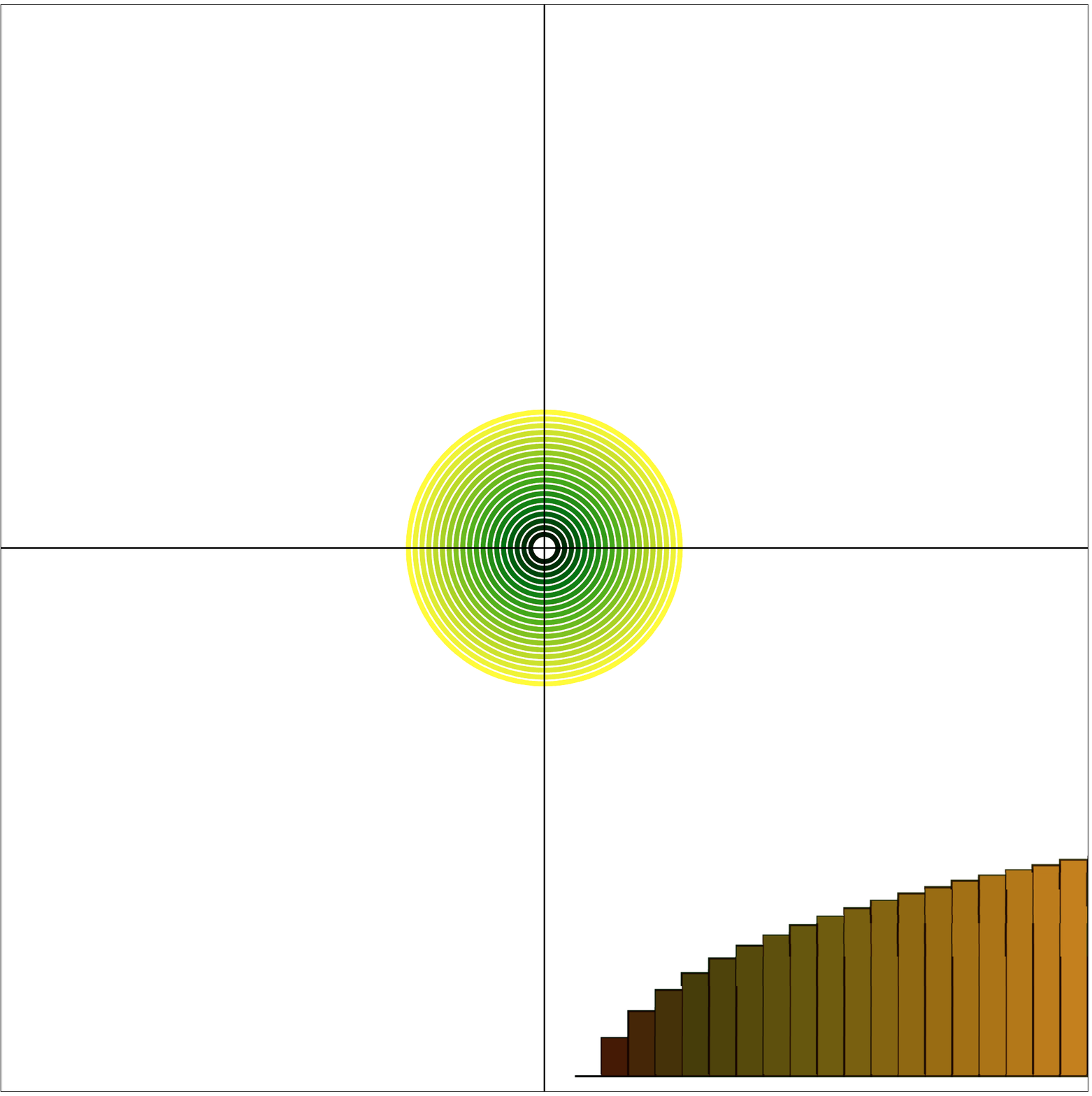}
\includegraphics[width=0.4\linewidth]{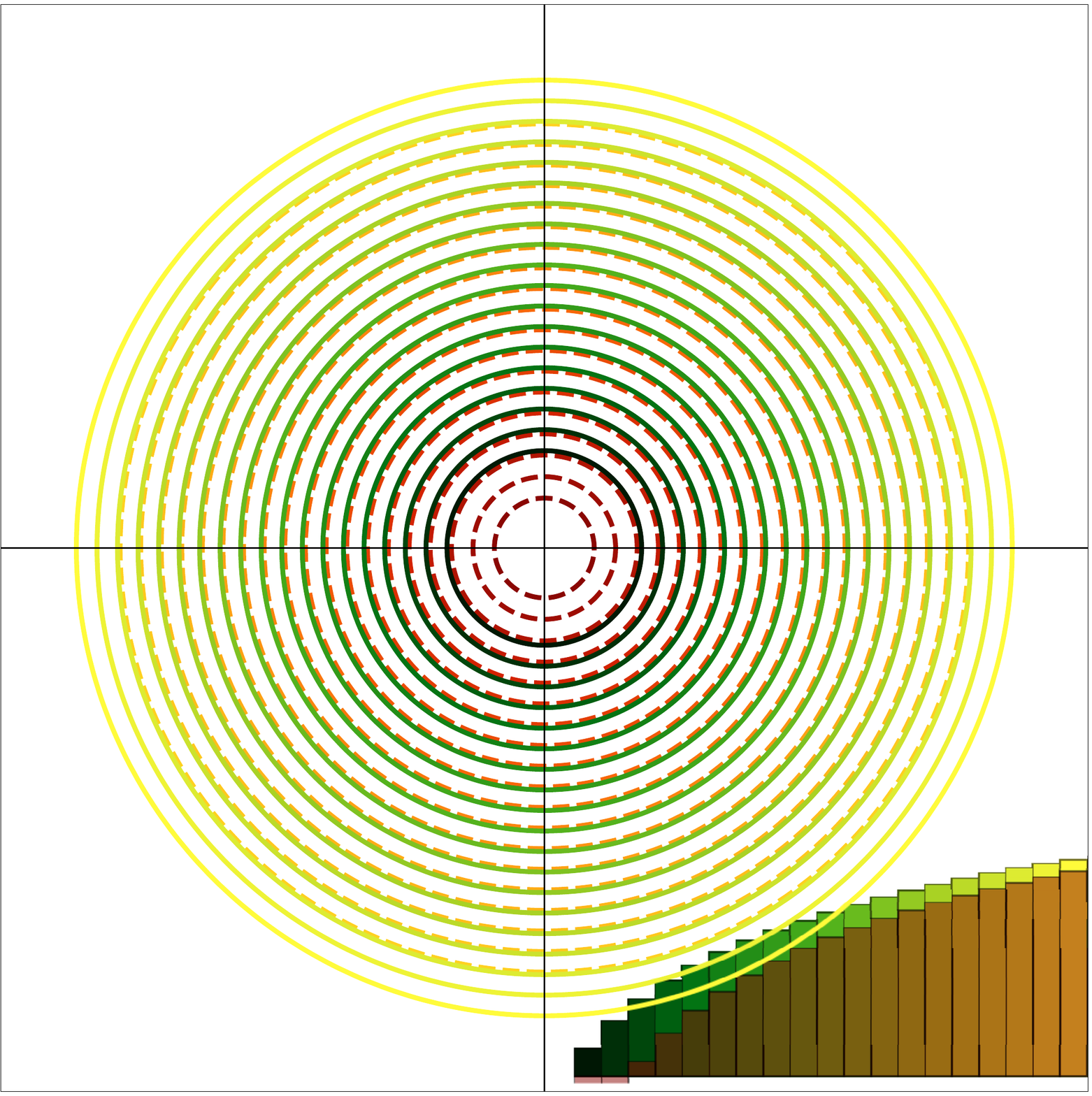}
   \caption{These snaptshots show, in arbitrary units, the numerical solution for the evolution of the shells without (top) and with (bottom) baryons together with the (log) mass distribution (insets), i.e. the mass inside the different shells from the inner most to the outer most. The upper snapshots show a case with shell-crossing and we can see how the mass distribution changes as the shells cross. Crucially, it is the very repulsive nature of the force what can lead to shell crossing in the expanding phase. This cannot happen for models with scalar fields due to their intrinsically attractive character. In the lower panels we can see how the charged DM shells separate from the baryons as they exit their screening radius.  Animations are available in the supplementary material.}\label{Fig:shells}
\end{figure*}

In the previous sections we have analysed the evolution of shells formed by identical particles. This would be the actual evolution if the dark electric force acted universally. We will now take a step forward and assume the perhaps more realistic situation where the initial distribution of particles contains both charged DM and uncharged baryons. This is a hypothesis which is similar to the one of coupled quintessence models where dark energy only couples to dark matter and is decoupled from baryons. We apply the same framework here to the dark electric force. We then have two different sets of equations: those for the evolution of baryons driven by gravity alone and the equations for DM that include the dark electric force. If we denote by $\Rdm(t,\rini)$ and $\Rb(t,\rini)$ the Eulerian coordinates of the DM and the baryons respectively, the system will evolve according to the following equations:
\begin{eqnarray}
\label{eq:animB1}
\ddot{R}_{\rm B}(t,\rini)&=&-\frac{GM(\Rb)}{\Rb^2}\,,\\
\label{eq:animB2}
\ddot{R}_{\rm DM}(t,\rini)&=&-\frac{GM(\Rdm)}{\Rdm^2}\Big[1-\beta^2 F(\Rdm/\rs)\Big]\,.
\end{eqnarray}
Notice that we are using the same Lagrangian coordinate for both components, which arises from assuming that they comove initially when only gravity acts.
We will use initial conditions analogous to those used for the pure DM distributions above, but now the initial shells have two components so we define the initial mass of the $i$-th shell as
$m_i=m_{i,\rm DM}+m_{i,\rm B}$. Since the baryons and DM shells evolve differently, now the mass of each initial shell is not conserved. The initial density profile is assumed to be uniform for both components so, regardless of its specific composition, the initial density is constant $\rho(\tini)=\rho_\star$. Thus, if we have a fraction $\Upsilon$ of baryons (that we also assume uniform so it is the same for all the shells), the mass of each shell is given by
\begin{eqnarray}
m_{i,\rm B}&=&\Upsilon\frac{4\pi\rho_\star}{3}\Big(R_i^3-R_{i-1}^3\Big)\Big\vert_{t=\tini}\,,\\
m_{i,\rm DM}&=&(1-\Upsilon)\frac{4\pi\rho_\star}{3}\Big(R_i^3-R_{i-1}^3\Big)\Big\vert_{t=\tini}\,.
\end{eqnarray}
Typical values of the initial baryonic fraction are $\Upsilon\simeq 0.2$ in accordance with the observed relative abundance of baryons and DM in the Universe. The total mass function $M(R)$ is then computed as in the single component case but keeping in mind that the sum extends to both baryons and DM, i.e. 
\be
M(R_i)=\sum_{R_{j,\rm B}\le R_i}m_{j,\rm B}+\sum_{R_{j,\rm DM}\le R_i}m_{j,\rm DM}\,.
\ee
Initially, when the electric force is screened, both components evolve together in comoving motion under the action of gravity. As the DM shells start exiting their screening radii, the electric force initiates the corresponding repulsion for the DM sector. At this point, the DM shells start expanding faster than the corresponding baryons shells. This causes a mass gain for the DM shells due to the baryon shells that are absorbed. The effect on baryons shells is however a mass loss due to the faster expansion of the DM shells. A crucial effect of this more involved evolution is that now we can have a shell crossing for the DM component induced by the baryons, while the baryon shells do not undergo shell crossing. In this case, besides the relative strength of the electric repulsion $\beta^2$ and the smoothness of the transition, the fraction of baryons $\Upsilon$ also plays a crucial role for the appearance of shell crossing since this parameter controls the mass gain of the DM shells when they expand faster than the baryonic shells.  If we look at the expansion rates of the baryons an DM shells we see that both are modified and acquire an inhomogeneous evolution in the transient phase. The effect is stronger for the DM component because it is this component that is affected by the electric force, while baryons are only affected by the mass loss. As we can see in the lower panels of Figure \ref{Fig:RBDM} (see also figure \ref{Fig:shells} bottom panel  and \cite{shells}), the variation on the DM shells is much stronger and follows a similar tendency to the case when no baryons are present, although the effects of the baryons is also apparent. For baryons we can clearly see a reduction in the expansion rate during the transient phase that is more pronounced for the outermost shells, but the effect is substantially smaller than for DM, as expected. Of course, the relative strength in the effects on the DM and the baryons depends on the corresponding ratio $\Upsilon$ that we have kept fixed to be small.

In the asymptotic phase we again recover the comoving evolution and the initially uniform distribution turns into an inhomogeneous profile. Again, the inhomogeneity is more pronounced for the DM component than for baryons. In fact, we can see in the upper right panel of Figure \ref{Fig:RBDM} that the DM shells become strongly packed into a small region, indicating a strong increase in the density for the DM component. It is noteworthy that both components reach the comoving motion but the inhomogeneous scale factor for baryons and DM will be different, i.e., we cannot globally describe the matter evolution with one single inhomogeneous scale factor.


\subsection{The $H_0$ tension}
\label{Sec:H0}

In the previous Sections we have seen how the expansion of the shells is modified by the presence of the electric repulsion. From the numerical solutions we have corroborated that the expansion rate of the outer shells is reduced with respect to the one of the inner shells when the screening ceases. At some point, the evolution for the DM shells is reversed, while for baryons the inner shells always expand slightly faster in the transient region for the considered values of the parameters. This slowing down of the outer shells can be understood from the effective Friedmann equation deduced above. Interestingly, although one may be tempted to think that the reduction in the expansion rate of the outer shells is driven by the effective Newton's constant, this is not the case and the dominant contribution comes from the inhomogeneous spatial curvature that the electric potential creates. In view of these results, it is easy to understand how this mechanism provides a promising scenario to alleviate the $H_0$ tension: The Hubble constant measured locally corresponds to the inner shells that exhibit a larger expansion rate due to the electric interaction as compared to the cosmological values that correspond to the outer shells.

It is convenient to consider this scenario in more detail to clarify some subtle points. A first point worth clarifying is the existence of two background metrics. In the relativistic view in terms of Lema\^itre models of Section \ref{Sec:Lemaitre} we were dealing with a single component Universe where the entire matter sector was universally coupled to the dark electrostatic interaction. An important consequence of this assumption is obviously that universality remains (i.e. a sort of {\it cosmological equivalence principle} still holds) so that we can describe the motion of particles in terms of a unique metric. In a more realistic scenario with uncharged baryons, this universality is broken\footnote{For the amusement of the reader enjoying semantic clarifications, let us stress that the (gravitational) equivalence principle is still valid. However, the presence of the electrostatic cosmological background affecting only the DM sector could be (to some extent mistakenly) interpreted as a cosmological violation of the equivalence principle.} and baryons and DM evolve according to different metrics (scale factors). Of course, the reason for this is that DM particles are subject to the long range electrostatic interaction while baryons are not. Thus, in this two-component Universe we have two metrics and the natural question that arises is: what metric would be probed by photons? The answer seems to be: neither and both. To explain why this is the case we need to bear in mind that photons would be emitted by galaxies that can be assumed to be inside virialised DM halos. In this scenario, galaxies are tracers of the DM distribution\footnote{We are assuming that galaxies are efficiently dragged by the gravitational potential of the DM halos, which is a reasonable assumption if the fraction of baryons is sufficiently small.} so that the emitting sources of photons follow the Hubble flow of the DM component. This would provide the initial condition to the photon's trajectories. However, if photons are not charged under the dark U(1) interaction, their propagation towards the receiver is oblivious to the direct  effect of the electric interaction. Thus, in their propagation they will probe the baryonic metric, which is the metric that drives the dynamics of the uncharged sector. We can be more quantitative by considering nearby objects for which we can obtain the Hubble law. In that case, the redshift of a photon emitted by a galaxy that belongs to the DM halo at position $R_{\rm DM}$ is given by $z=\dot{R}_{\rm DM}\simeq H_0 R_{\rm DM}$, i.e., the redshift can be fully ascribed to the recession velocity of the galaxy so we will be probing the local value of $H_0$ of the corresponding shell. Since these nearby objects live in the inner shells, the measured value will be higher than the cosmological one, inferred from CMB for instance, that would correspond to the outer shells. 

\begin{figure*}
\includegraphics[width=0.49\linewidth]{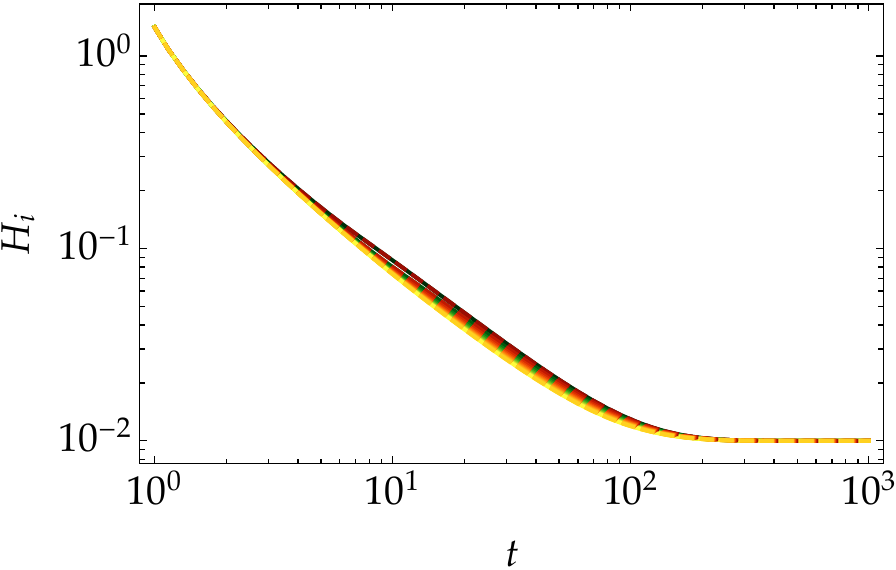}
\includegraphics[width=0.49\linewidth]{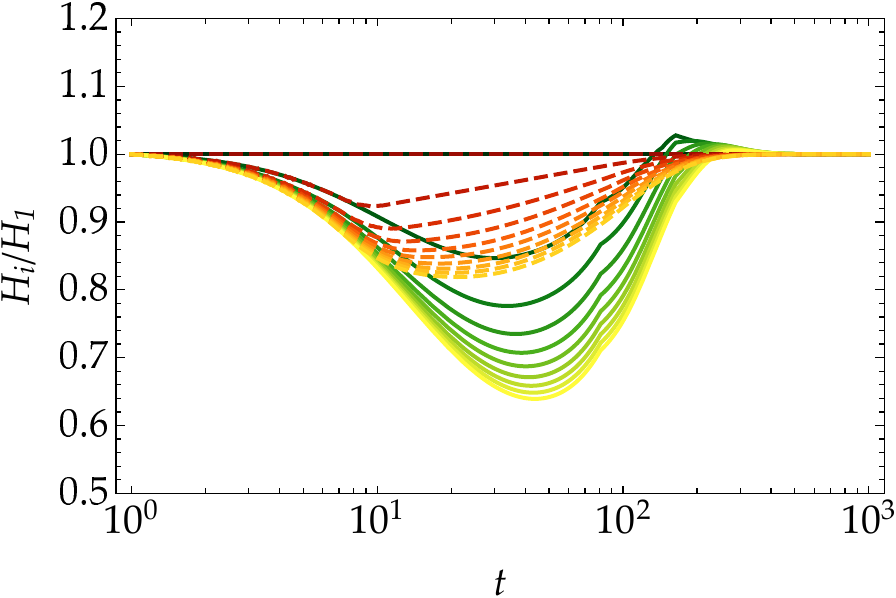}
\caption{This figure shows the evolution of the expansion rates for the shells with the same parameters as in Fig. \ref{Fig:riBI} and in the presence of a cosmological constant with $\frac{8\pi G\rho_{\Lambda}}{3}=10^{-4}$. Notice that the inner shells have a larger expansion rate than the outer shells. The figure on the left corresponds to the absolute expansion rates for all shells and the convergence to a Universe dominated by a cosmological constant can be seen. The right figure gives the normalised Hubble rate to the inner most shell. Baryons are represented in red whilst dark matter is in green. }
\label{Fig:shellsDE}
\end{figure*}

So far, we have only considered the dust component and a fair objection could be that the present Universe is dominated by dark energy which could play an important role. It is not difficult to include a cosmological constant in our numerical set-up. Since a cosmological constant has a constant density $\rhode$, its effect can be easily accounted for by adding an uncharged mass $\Delta M= \frac{4\pi R_i^3}{3}\rhode$  to the $i-$th shell, i.e., adding a term proportional to $R_i$ to the rhs of the evolution equations for both DM and baryons:
\begin{widetext}
\bea
\ddot{R}_{\rm B}(t,\rini)&=&-\frac{GM(\Rb)}{\Rb^2}+\frac{4\pi G\rhode}{3}\Rb\,,\\
\ddot{R}_{\rm DM}(t,\rini)&=&-\frac{GM(\Rdm)}{\Rdm^2}\Big[1-\beta^2 F(\Rdm/\rs)\Big]+\frac{4\pi G\rhode}{3}\Rdm\,.
\eea
\end{widetext}
Notice too that the inclusion of dark energy has already been discussed in the case of the Lema\^itre models \ref{Sec:Lemaitre}.
With the inclusion of this term, we obtain the expansion rates depicted in Figure \ref{Fig:shellsDE}. The obtained results can be easily understood in view of the modified Friedmann equation. Since the inhomogeneous curvature contribution from the electric force is larger for the inner shells, their accelerated expansion caused by the cosmological constant is effectively delayed with respect to the outer shells. In other words, the breaking of comoving motion induced by the electric force makes the different shells to enter the accelerated regime at different times. Consequently, the local Hubble factor as measured from the inner shells will be slightly larger than the one corresponding to the outer shells. As this is what seems to be the observed results between large and local expansion rates, we expect that the detailed analysis of the dark models presented here could be made to reproduce current data. This is left for future work.

 Nonetheless we can give an approximate description of the way the local value of $H_0$ is modified following Appendix \ref{sec:simpl} where we have discussed the Friedmann equation and its consequences in a simplified context where dark matter and a cosmological constant are taken into account.
 We find that when shell crossing does not happens, as for the Born-Infeld model, the local Hubble rate is related to the large scale one by 
\begin{equation}
    H_0^{\rm local}\simeq H_0^{\rm CMB}\left(1+ \frac{\beta^2}{2} \Omega_{m0} z_s\right)
    \label{eq:HH}
\end{equation}
where $z_s$ is the redshift of exit from the screening radius of a shell labelled by $r$ and $\Omega_{m0}\sim 0.25$ is the dark matter fraction. First of all notice that the local Hubble rate is always larger than the large scale one as long as $z_s>0$. This is guaranteed as long as the horizon becomes unscreened in the past of our Universe, see section \ref{Sec:Discussion}. Indeed,  as  $a_s\propto \sqrt{r}$ we find that inner shells are always unscreened earlier than the whole horizon. For the same reason shells with differing $r$ have different $z_s$, i.e. local Hubble rates measured using probes on different scales $r$ will have different Hubble rates. In order to be compatible with observations, one must also make sure that  the BAO occur in the unscreened regime. This can be achieved if the whole horizon is in the screened regime until a redshift of order $0.5$. This guarantees that all fluctuations within the horizon feel a suppressed dark force due to the large ${\cal K}_Y$ factor in (\ref{supp}). In section \ref{Sec:Discussion}, we will see that for $\beta={\cal O}(1)$, and 
$10^{-3}\ {\rm eV} \le \Lambda_e \le 10^{-2} {\rm eV}$, the horizon becomes unscreened for $z\le 0.5$ and galaxies are typically unscreened too. As the inner shells become unscreened earlier than the whole horizon,  we can  take as a template $z_s\sim 1$ and  $\beta={\cal O}(1)$.  Using these values and (\ref{eq:HH}), the resulting deviation of the local Hubble rate from the large scale one is then around ten per cent as measurements seem to indicate. Of course, in order to address the Hubble tension and to be compatible with BAO being screened, the transition between the two regimes needs to be sufficiently fast. This will ultimately depend on the specific function of the non-linear electromagnetism. A more thorough description of this effect is left for future work.

As already mentioned, this suggested mechanism to alleviate the $H_0$ tension differs from other proposals relying on scalar fields (see e.g. \cite{Knox:2019rjx,Zumalacarregui:2020cjh,Ballesteros:2020sik,Braglia:2020iik,Desmond:2020wep} but also \cite{Mortsell:2018mfj,Dhawan:2020xmp} for a more phenomenological approach) in at least two ways. Firstly, the main effect to alleviate the $H_0$ tension is through a change in the effective Newton's constant, which is not what we do here. In some of those models  the early background cosmology is affected \cite{Bernal:2016gxb,Poulin:2018cxd,Agrawal:2019lmo,Alexander:2019rsc,Lin:2019qug}\footnote{It has been recently noticed that, despite early DE models could  alleviate the $H_0$ tension, they would, at the same time, worsen the $\sigma_8$ tension \cite{Hill:2020osr}.}. In our scenario however, even though there is also a modification of the effective Newton's constant (with interesting phenomenological consequences that we will discuss below), the mechanism relies on the screening mechanism that unleashes a late time repulsive force that breaks the comoving motion, so the expansion becomes inhomogeneous, and makes the local Universe expansion rate stronger than the cosmological one. Let us stress once again that this mechanism crucially depends on the spin-1 nature of the screened interaction and a similar scenario for scalar fields is not possible due to their attractive nature (at least without invoking contrived interactions). 

It has recently been suggested \cite{Kazantzidis:2020tko} by a tomographic analysis of the Pantheon supernovae data set that the local Universe could indeed have a value of $H_0^{\rm local}$ of about $2\%$ larger than the cosmological value $H_0^{\rm cos}$. In that study, the authors interpreted the result as an indication of a local underdensity. An earlier analysis of the Pantheon dataset found similar results  \cite{Colgain:2019pck}. It is remarkable that our scenario could indeed explain this result in a natural and theoretically motivated manner since that is precisely the obtained result. The larger value of $H_0^{\rm local}$ arises from the faster expansion of the inner shells.


\section{Some phenomenological aspects}\label{Sec:Discussion}

In the previous section we have seen how the present model could ease the $H_0$ tension and ideally reconcile local measurements with the ones from CMB. However, the expected phenomenology associated to a charged DM is far richer. We will devote this section to a brief discussion of the aspects that we believe are most interesting. 
 
\subsubsection{Some more on the cosmological evolution}

From a cosmological perspective, the physics which takes place before the time when the screening radius becomes smaller than the horizon is unchanged, as can be see from the evolution of matter shells in figure \ref{Fig:RBDM}.  This is a direct consequence of the fact that the ratio between the two aforementioned scales goes as
\be 
r_s^{(H)}(a)H(a) \propto a^{-p/4}
\ee 
where $p = 3(1+w)$ and  $w$ is the equation of state parameter of the matter species dominating the Universe. Since this ratio is a decreasing function of the scale factor\footnote{{It is interesting to notice that if a cosmological constant term is dominating, the ratio becomes constant. Hence, if the transition to the unscreened cosmological regime does not occur prior to cosmological constant domination, it will never occur.}}, it is quite natural to assume that in the early Universe the DM dynamics is blind to the force. In particular, we require that at last scattering the whole horizon is inside its screening radius in order  to avoid sizeable modifications to the CMB physics. For $\beta={\cal O}(1)$, this results in an upper bound on the energy scale $\Lambdae$, namely
\be 
\Lambdae \lesssim \mathcal{O}(1)\ {\rm eV}\,.
\ee
Notice that as long as the whole horizon is screened any structure or fluctuation of the matter density is screened and is therefore blind to the presence of the dark force. This follows from suppression of the dark repulsion by the large factor ${\cal K}_Y$ as in (\ref{supp}). This effect is analogous to the same phenomenon for scalar K-mouflage models \cite{Brax:2014wla,Brax:2014yla}.
On the other hand, to have the force active at recent cosmological times we require that the screening scale today is smaller than the size of the horizon. This in turns implies a lower bound on the energy scale $\Lambdae$, for $\beta={\cal O}(1)$, of the order
\be
\Lambdae \gtrsim 
10^{-3}\ {\rm eV}\,.
\ee
We conclude that the energy scale for the non linearities needs to be in the range
\be\label{eq:Lambda_range}
10^{-3}\ {\rm eV}\lesssim \Lambdae\lesssim 1\ {\rm eV}\,.
\ee
Notice that for $\Lambdae = 10^{-2}$ eV we get that the horizon  equates its associated screening radius at $z\sim 0.6$ while for $\Lambdae = 10^{-1}$ eV the equivalence occurs at $z\sim 30$.

It is worth notice that, even though at last scattering the electrostatic force is absent, there can be effects on the CMB spectrum. For example, since the DM distribution will be modified as compared to the uncharged case, we can expect to see differences accumulating as the CMB photons travel to us affecting, for example,  the late-time Integrated Sachs-Wolfe effect. 

\subsubsection{Astrophysical aspects}

Once the horizon scale becomes larger than its screening radius, the electrostatic force switches on leaving a redshift and scale dependent modification to the DM Hubble law \eqref{Eq:Friedmann2}. As we have seen in  section \ref{Sec:H0}, this results in a larger local  expansion rate as compared to the one on larger scales. 
However, this is just one side of the effect of the electrostatic force. In fact, once we are in the cosmological unscreened regime we need to analyse if and when virialized objects  actually feel the repulsive force. In other words we need to compute the screening scale associated to the mass enclosed by a certain DM object. Since we are considering a late time transition to the unscreened cosmological regime, we will assume that the baryonic inhomomogeneities will trace the DM distribution and, particular, we will consider galaxies to be dragged by DM halos\footnote{It is interesting to notice that due to the repulsive nature of the electrostatic force, DM-less galaxies could, in principle, form.}.

Let us focus on the expected values of the screening radius for clusters and galaxies.  Taking for the typical size of clusters $5$ Mpc and masses of order $10^{15} M_\odot$,  we find that a cluster is screened if the cut-off scale  is
\be 
\Lambdae \lesssim 10^{-4}\ {\rm eV}\,.
\ee
Taking into account the cosmological bound \eqref{eq:Lambda_range} we see that in general clusters will not be screened today.
For galaxies we take a typical size of $100$ kpc and masses of order $10^{12} M_\odot$ and we get that galaxies are screened if
\be 
\Lambdae \lesssim 10^{-3}\ {\rm eV}\,.
\ee
Hence, we see that galaxies as well will, generically, not be screened. However, depending on the specific properties and on the detailed cosmological evolution certain types might be screened. We can then conclude that, as soon as the horizon becomes larger than its associated screening radius, most of the astrophysical objects will start to feel the repulsive force. In this regime, the electrostatic interaction will start competing with the gravitational interaction and its main effect is a redressing of Newton's constant as 
\be
G_{\rm eff} = G (1-\beta^2)
\label{newG}
\ee 
effectively reducing the gravitational interaction.
 
 Let us first elaborate more on how the electrostatic force can impact the determination, for example, of the dynamics of standard candles and their calibrators. In the previous section we have seen how photons' propagation,  will be modified by the presence of the electrostatic force only indirectly via the gravitational potential of DM structures along the path. Hence, the modification of the Hubble law can impact the determination of the luminosity distance for objects that are at sufficiently high redshift. We will focus now on a supernovae belonging to a nearby, fully virialised galaxy. Then the supernovae is subject to the peculiar velocity of the host galaxy.
When its distance to our galaxy is smaller than twice the screening radius $r_s$, the electrostatic interactions is negligible.
On the other hand, when it is further away than $2 r_s$, the electrostatic interaction will start to act as a new force
\be
\vec F_{12}= \frac{\beta G   m_1 m_2}{ \vert \vec x_1 -\vec x_2 \vert^3} (\vec x_2 -\vec x_1)\,.
\ee
This implies that the determination of the relative motion between the host and our galaxy  will be modified due to the repulsion in a space dependent manner. Moreover, if the horizon crosses the cosmological screening radius at very recent times, we can also expect a redshift dependence. In fact, the light from a high redshift supernova might have been emitted while the electrostatic force was cosmologically screened even if the host galaxy is not  massive enough to be self-screened.
Let us now consider an ensemble of galaxies in a cluster. 
Since neither the cluster nor the galaxies belonging to it are generically screened, there will be two main effects on the dynamics. Firstly, the bulk motion of the cluster will be affected because of its total dark charge, analogously to what we described for above for galaxies.  Secondly, each galaxy will be subject to the repulsive force generated by all the others.
Assuming virial equilibrium, the total mass of the cluster as derived from the motion of galaxies can be obtained via \cite{Carlberg:1995ra,Lokas:2003ks}
\be\label{eq:Mcluster}
\frac{G_{\rm eff} M}{2 R}\sim \frac{3}{2}\sigma_r^2
\ee 
with $\sigma_r$ the observed galaxies'  velocity dispersion\footnote{More precisely, the observable quantity is the line of sight projected  velocity  dispersion.}, $G_{\rm eff}$ is the effective Newton constant experienced by galaxies and $R$ is the virial radius of the cluster.
On the other hand, the total mass of the cluster can be inferred via other independent measurements. Even more interesting is the fact that such measurements, at least at first order in Newtonian expansion, are not affected by the electrostatic force. For example, the total mass can be reconstructed by looking at the intracluster gas distribution that represents the dominant baryonic mass component of a cluster \cite{Reiprich:2001zv}. Since baryons are not charged they will be sensitive to $G $.
Another way of getting the mass is to use gravitational lensing \cite{Hoekstra:2013via}\footnote{In fact, gravitational lensing is sensitive to the lensing potential which, to leading order and in the absence of anisotropic stresses or, equivalently, with a trivial slip parameter, is given by twice the Newtonian potential. This allows one using the Poisson equation depending on $G $ to reconstruct the density field.}.
Hence, by knowing the total mass and the cluster's radius from independent measurements we can use \eqref{eq:Mcluster} to place a constraint on the value of $\beta^2$ as
\be 
 \beta^2 \le \left\vert 1-\frac{\sigma_r^2}{(\sigma^{(N)}_r)^2}\right\vert 
\ee 
where $\sigma^{(N)}_r$ is the dispersion velocity inferred (indirectly) using Newton's constant. If we assume that the correction is small we get
\be 
\beta^2 \le  \frac{\vert\delta\sigma_r\vert}{\sigma_r^{(N)}}\,.
\ee 
Of course, this is just an extremely rough estimate of the size of the corrections as we are neglecting several important contributions, both astrophysical and model related. For example, there is no  reason a priory to expect a standard DM profile for the cluster. Also, assuming the same lensing potential as in GR is only an approximation as the two gravitational potentials are not generically equivalent.  Moreover, the different background evolution should also be taken into account when estimating the distance of the cluster. All in all, the goodness of the available data  \cite{2012ApJS..199...25P,Biviano:2013eia,Caminha:2019veg} and the compatibility of mass measures from different tracers \cite{Donahue:2014qda} make galaxy clusters an ideal astrophysical laboratory to test the presence of an electrostatic force in the DM sector.

Another interesting consequence that can be drawn is that, due to the repulsive nature of the interaction, fewer small mass DM halos are expected to form in this model. This suppression of the DM mass function could ease the tension between the predicted number of satellite galaxies and the one actually observed \cite{Klypin:1999uc,Bullock:2017xww}, although this mismatch can find  an explanation also in the context of uncharged DM \cite{Kim:2017iwr} or through baryonic physics \cite{Brooks:2012ah}. We could also expect that the repulsive force may deplete some regions of space and create voids. For all these phenomena related to the formation of structures, we expect that numerical simulations with N-body codes including the dark force will certainly provide a more thorough picture than the one we have used here with spherical shells. These simulations are in the process of being performed. On the other hand, we expect that the description using shells being non-linear does capture the essence of the background dynamics of the Universe even down to non-linear scales. In particular as most of structure formation occurs at redshifts $z\gtrsim 0.5$ where the horizon is screened and the growth of structure is, as a result, not modified, we expect that effects of the dark force on structure formation will be bounded. Constraints of the type obtained in \cite{Kesden:2006vz}\footnote{This paper focuses mainly on attractive scalar interactions between dark matter particles and no such interactions between baryons. The case of repulsive scalar forces is  also briefly discussed corresponding to a change of $G$ as in (\ref{newG}). } from tidal disruption of satellite galaxies are expected to apply and lead to bounds on $\beta$ at the $0.1$ level. In particular, the fact that baryons are not charged whilst dark matter feels the dark repulsion in unscreened galaxies would play a prominent role and would lead to an effective violation of the equivalence principle for large $\beta$. The analysis of these phenomena would certainly require new  numerical simulations. This is left for future work.

Furthermore, since in our model baryons are uncharged, their distribution will have a larger bias as compared to the case of uncharged DM. An extreme situation is represented by clusters collisions where we expect to see  DM to show more resistance in crossing through than in standard cases. In particular, we would expect to see a mismatch in the position of the centres of mass of DM and galaxies (that can be considered effectively collisionless, contrarily to the gas component of the cluster). 

Finally, let us comment on potential constraints imposed by a displacement of the Milky Way from the center of the charged dark matter distribution. When it comes to the CMB the main potential conflict comes from the CMB dipole which, however, is not affected within this scenario. In fact, the local effect would be a contribution to our local motion, but this would only correct our peculiar motion with respect to the CMB. Since it is a genuine Doppler effect (amplified by the extra force) the constraints from aberration from Planck are evaded \cite{Aghanim:2013suk} (see also the recent \cite{Ferreira:2020aqa}). Possibly, the most relevant effect would be the observation of an anisotropic modulation in the Hubble diagram. Although this has been constrained by several analyses with even claims of statistically mild detections (see e.g. \cite{Schwarz:2007wf,Colin:2010ds,Jimenez:2014jma,Bengaly:2015nwa}), these are much less constraining than the CMB dipole ones. This point is discussed in more detail in \cite{BeltranJimenez:2021imo} where it is shown that there is no strong angular dependence provided the sources are further away from us than the center.

\subsubsection{Charging baryons}\label{sec:chargedbaryons}

Let us finally mention the intriguing possibility to charge also the baryonic sector. In the early Universe the cosmological screening would apply and its dynamical behaviour will be as in the uncharged case, exactly as it happens for DM during that stage. It is only in the recent Universe, when the screening radius enters the horizon, that the new physics kicks in. Since, in general, galaxies will not be screened we expect to have new interesting physics at both the DM and baryonic level. On the other hand, solar system and laboratory tests strongly constrain any fifth force acting on Standard Model particles. Hence, the first requirement is to have at least the solar system safely screened. Indeed, the screening radius of the sun is
\be
r_{s,\odot}
\sim  \lambda_b\left(\frac{{\rm eV}}{\Lambdae}\right)r_{\rm \odot N}
\ee
where $r_{\odot N}$ is the semi-major axis of Neptune orbit and $\lambda_b$ is {\bf related to the baryonic dark coupling $\beta_b$. } Hence, we see that if the solar system needs to be screened, then either the constraint on $\Lambdae$ tightens or baryonic matter has a much larger charge compared to DM. 
Of course, the latter shows that the case where baryons are uncharged is peculiar. When uncharged the solar system constraints become void. On the other hand, if a charge is present for the baryons then the coupling to the dark force must be large enough. 

We move now to  discuss  briefly how laboratory experiments could cast constraints on this kind of interactions. 
When baryons are coupled we expect a modification of Newton's law on scales much larger than the screening length. Typically, the screening length for a coupling $\beta_b$ of order unity is given by
\be 
\rs \simeq \sqrt{\frac{m}{m_{\rm Pl}}} \Lambdae^{-1}.
\ee
For test masses of order $1$g we get screening radii of order $\rs\simeq 10^3 \Lambdae^{-1}$. Taking for $\Lambdae$ values between the one tenth of the dark energy scale and 1 eV corresponding to the values for which the whole Universe is screened at last scattering and galaxies are screened now, we find screening radii between one metre and $0.1$ mm. The existence of Newtonian forces on distances larger than $0.1$ mm has been tested by the E\"otwash experiment \cite{Kapner:2006si}. Here a new analysis would have to be performed taking into account the screening of charges and the presence of a shield for electrostatic interactions. A better prospect may come from atomic interferometry \cite{Hamilton:2015zga,Elder:2016yxm,Sabulsky:2018jma} where a large ball of Aluminium influences the behaviour of Caesium atoms at a distance of 2 cm. For balls of radius about 1 cm and masses around 10 grams, we find that the screening is larger than 2 cm when typically we have $\Lambdae \le 0.1$ eV. Of course a proper analysis should be devoted to constraints coming from laboratory experiments. This is left for future work. 

\section{Conclusion}

The growing tension between early (CMB) Universe observations and local measurements in our galactic environment is beginning to shake our understanding of both its cosmological evolution and  matter content. In particular, the mismatch between the value of $H_0$ as inferred from the Planck satellite \cite{Aghanim:2018eyx} and as measured from Supernovae (and other local measurements) \cite{Riess:2019cxk,Wong:2019kwg,Birrer:2018vtm} may call for an overhaul of the standard model of cosmology, i.e. the $\Lambda$-CDM description where baryons, dark matter, and dark energy are the main components of our Universe with universal interactions governed by gravity. 

Motivated by these observational tensions, in this work we have explored the possibility of extending the arsenal of fundamental forces acting on very large scales and its potential observational effects. One of the features of gravity is the observed absence of negative masses and its universal attractiveness.
Although other fundamental interactions such as electromagnetism share the long-range character of gravity, their action is screened on very large scales owed to the neutral balance between positive and negative charges for astrophysical objects. 

In this paper, we have analysed the role that
an additional electromagnetic interaction, dark electromagnetism, could play on large scales. Large distance effects on the dynamics of the Universe are guaranteed when the dark charges of matter under this new U(1) field are all of the same sign, mimicking what happens for gravity.
On very large scales, the repulsiveness of the interaction between matter particles counteract the gravitational pull and could have consequences on the cosmological background evolution and the dynamics of galaxies and clusters. On the other hand, a large repulsion between matter objects is certainly prohibited by the absence of deviations from gravity in the solar system and the successes of the description of early Universe cosmology up to the last scattering time and the acoustic oscillations of the CMB. These successes can be preserved if the new electromagnetic interaction is screened due to its non-linear character on short distance scales. This feature would preserve the description of the early Universe including the CMB if the screening radius is larger than the horizon until a redshift less a few hundreds. 

Here we have investigated the case of dark matter being charged under this dark and non-linear U(1) gauge interaction. The non-linearities in the dark electromagnetic sector are the very distinctive property of our scenario as compared to other models with charged dark matter and additional dark long-range interactions. We have  explored its dynamics in the context of Newtonian cosmology as a fully relativistic treatment is fraught with ambiguities due to the long range nature of the new interaction. If the early Universe underwent a phase of dark matter genesis in which only one type of charge survived, the Universe would be filled with such an interacting DM component. The non linear nature of the dark electromagnetic force splits the dynamics of the Universe into two distinct regimes characterised by the screening radius $r_s$ defined in \eqref{eq:r_s}. At separations smaller than the screening scale the interaction is suppressed making the dynamics of DM indistinguishable from that of the uncharged case, while at larger separations DM particles start to feel the repulsion due to the interaction. As we have seen in section \ref{Sec:Lemaitre} this electrostatic force has the geometrical interpretation of particles moving in an inhomogenous spherically symmetric Universe. On the other hand, particles that are uncharged under the U(1) interaction will follow geodesics associated to an uncharged metric. This does not imply that the two fluids evolve independently  as we have seen in section \ref{Sec:numerics}. The uncharged sector  feels the presence of the dark force which constrains the Universe to become inhomogeneous as the different Newtonian shells, i.e. the different spherical shells labelled by the initial comoving radii in the early Universe,  become unscreened at different times. This has important consequences for the late time dynamics of the Universe. In particular we have shown that the innermost shells, i.e. objects in our local environment, would become unscreened earlier than outermost shells corresponding to more distant objects. As a result, not only the Hubble rate of nearby objects would be larger than in the early Universe, mimicking the observed discrepancy between local and CMB data, but local measurements of the Hubble rate would differ between close objects and further ones, e.g. implying a different Hubble rate for local supernovae and cosmological ones. Of course we have not yet carried out a full quantitative analysis of this phenomenon and this can only be considered so far as a scenario. More thorough studies are left for future work (see however \cite{BeltranJimenez:2021imo} for preliminary work in this direction).

The existence of the dark repulsion could also be traced in the dynamics of galaxies and clusters. Indeed it turns out that they are unscreened almost as soon as the screening radius enters the horizon and as such would feel the extra repulsion. This could have observed effects in the peculiar velocities if they are reconstructed assuming the Newtonian dynamics of gravity. This would follow from the reduction of Newton's constant induced by the repulsive interaction on unscreened objects. Similarly the collapsing dynamics of spherical shells should also be affected implying plausible consequences for large scale structure formation and cluster number counts. 

Finally in this paper we have focused on the case where only dark matter could be charged under the new U(1) interaction. Many other possibilities could be envisaged. Baryons could be charged with consequences from large scale structures to laboratory experiments. Neutrinos could be charged with consequences on their time delays with photons. Of course we believe that large scale simulations of the dynamics of the Universe with this new electromagnetic interaction should reveal intricacies such that new effects for voids in the Universe. There can also be important consequences for astrophysical probes of dark matter annihilation/scattering. Although Sommerfeld enhancement is naturally suppressed inside big dark matter haloes due to the non-linear screening of the effective coupling constant, it can become relevant around low mass haloes. In that case however, the low DM density would play against it. On the other hand, at an even more speculative level, if compact objects carrying a non-trivial dark charge exist in unscreened environments, they could provide a population of {\it exotic} objects which could give detectable signals in gravitational waves. Around such compact objects where the dark electric field could be large, there would also be the possibility of producing dark matter particles via the Schwinger mechanism. In summary, the presented scenario constitutes a promising framework with interesting observational signatures in a wide variety of contexts. These applications will be explored in more detail in future work.

\textbf{Acknowledgments:} We would like to thank Marcello Musso for useful discussions. P.B. acknowledges relevant discussions with Alain Blanchard on the $H_0$ tension and BAO. JBJ acknowledges support from the  \textit{Atracci\'on del Talento Cient\'ifico en Salamanca} programme and the MINECO's projects PGC2018-096038-B-I00 and FIS2016-78859-P (AEI/FEDER).  This article is based upon work from COST Action CA15117, supported by COST (European Cooperation in Science and Technology).   DB  acknowledges  support  from the \textit{Atracci\'on del Talento Cient\'ifico en Salamanca} programme and the project PGC2018-096038-B-I00 by Spanish Ministerio de Ciencia, Innovaci\'on y Universidades. This project has received funding /support from the European Union’s Horizon 2020 research and innovation programme under the Marie Skłodowska -Curie grant agreement No 860881-HIDDeN.

\appendix

\section{Some theorems}
\label{app:A}

In this appendix we extend some well known theorems valid for linear electrostatic forces to the case in which non-linear corrections appear. This can be particularly relevant for laboratory experiments, in the case baryonic matter is also charged as we mentioned in section \ref{sec:chargedbaryons} .

\subsection{The cavity theorem}

Let us assume that the particles are evenly distributed  inside a spherical cavity of centre the origin of coordinates and radius $R$. Inside the cavity the field $E_r$ is spherically distributed and must satisfy
\bea
\vec \nabla\cdot( {\cal K}_Y \vec E)=0, \ \ r\le R
\eea
where $\vec E$ depends on $r$ and is radial. As a result ${\cal K}_Y$ only depends on $r$ too. Integrating this equality over a ball of radius $r$ and centered at the origin gives using Green's theorem that
\be
E_r(r)=0, \ \ r\le R
\ee
i.e. the electric field vanishes. Hence no effects of the particles outside the cavity are present inside the cavity.

\subsection{The effacement theorem}

Let us now consider the effects of the particles inside the cavity when no particles are outside. Again we must solve
\bea
\vec \nabla\cdot( {\cal K}_Y \vec E)=\rho_q \theta(R-r),
\eea
where the field $\vec E$ is radial and depends only of the radius thanks to the homogeneity and isotropy of the coarsed-grained distribution of particles.
Green's theorem tells us that
\be
{\cal K} _Y E_r = \frac{ \beta M(R)}{4\sqrt{2}\pi r^2}
\ee
i.e. the electric field is the one obtained by putting all the charge $\beta M(R)/\sqrt 2 m_{\rm Pl}$ at the origin. Here  $M(R)$ is the mass of the particles inside the ball.

\section{A simplified treatment of the $H_0$ tension}
\label{sec:simpl}

In this appendix, we present a simplified treatment, a  gedanken analysis, of the $H_0$ tension when only one species is present, i.e. dark matter, and is charged under the dark U(1). We also assume that no shell crossing happens as it is the case in Born-Infeld theory for instance. In the following, we take as a simplification that the function $F$ which governs the 
 transition between the screened to the unscreened regimes is sharp, i.e. $F=1$ in the unscreened region and $F=0$ when screening takes place. This implies that the Hubble rate when all scales feel the new interaction in an unscreened way reads
\be  
H^2= \frac{8\pi G }{3} \left[ \rho_{0}\left(\frac{a_0^3}{a^3} (1-\beta^2) +\beta^2\frac{a_0}{a_s}\frac{a_0^2}{a^2}\right) +\rho_{\rm DE}\right]
\ee
where $\rho_0= \rho_\star \frac{a_\star^3}{a_0^3}$ and $a_0$ is the scale factor of the observer at late time. For local objects we have $a\simeq a_0$, and their redshift provides a measure of their velocity. We assume that objects close-by  emit light whilst being in the Hubble flow of dark matter.  The local Hubble rate, i.e. the one of dark matter which corresponds to the Hubble rate at emission, is given by
\begin{widetext}
\be 
H_0^{\rm local}= H_0^{\rm CMB}\left(1 -\frac{\beta^2}{2} \Omega_{m0}+ \frac{\beta^2}{2} \Omega_{m0} \frac{a_0}{a_s}\right)= H_0^{\rm CMB}\left(1+ \frac{\beta^2}{2}\Omega_{m0}z_s\right)
\label{loca}
\ee
\end{widetext}
where $1+z_s= a_0/a_s$ and the Hubble rate $H_0^{\rm CMB}$ has been normalised in the absence of dark interaction, as befitting what happens in the early Universe when the screening radius is larger than the horizon, 
\be 
\Omega_{m0} H_0^2 =  \frac{8\pi G }{3}\rho_{0},\ \  \Omega_{\Lambda} H_0^2 =  \frac{8\pi G }{3}\rho_{\rm DE}
\ee
and $\Omega_{m0}+\Omega_{\rm DE}=1$. We can see the effects of the dark electric interaction in (\ref{loca}). The reducing of Newton's constant $-\frac{\beta^2}{2} \Omega_{m0}$ due to the repulsiveness of the electric force is largely compensated by the increase coming from the negative curvature effect due to  the electric pressure $\frac{\beta^2}{2} \Omega_{m0} \frac{a_0}{a_s}$. As a result, the local value of $H_0^{\rm local}$ is larger than the value obtained with the CMB normalisation. Moreover as $a_s$ is smaller for innermost shells, i.e. close objects, we can see that these emitting  objects have a larger Hubble rate than distant ones.

\bibliography{H0tension}

\end{document}